\documentclass[a4paper,11pt]{article}
\usepackage{jheppub} 
\usepackage[utf8]{inputenc}
\usepackage{subfig}
\usepackage{comment}
\usepackage{feynmp}

\usepackage{slashed}
\usepackage{tikz}

\newcommand{\gs}{g_\star}
\newcommand{\gss}{g_{\star s}}
\newcommand{\Trh}{T_\text{rh}}
\newcommand{\arh}{a_\text{rh}}
\newcommand{\Tmax}{T_\text{max}}
\newcommand{\amax}{a_\text{max}}
\newcommand{\aend}{a_\text{end}}
\newcommand{\rGW}{\rho_\text{GW}}
\newcommand{\rR}{\rho_R}
\newcommand{\rp}{\rho_\phi}
\newcommand{\oGW}{\Omega_\text{GW}}

\newcommand*\circled[1]{\tikz[baseline=(char.base)]{
		\node[shape=circle,draw,inner sep=0.5pt] (char) {#1};}}

\newcommand{\Eom}{E_\omega}
\newcommand{\Hinf}{H_{\text{inf}}}

\title{\Large Ultra-high Frequency Gravitational Waves \\from Scattering,  Bremsstrahlung and Decay during Reheating}

\author[]{Yong Xu}
\affiliation[]{\it PRISMA$^+$ Cluster of Excellence and Mainz Institute for Theoretical Physics\\
	Johannes Gutenberg University, 55099 Mainz, Germany}
\emailAdd{yonxu@uni-mainz.de}

\abstract{We investigate ultra-high frequency gravitational waves (GWs) from gravitons generated during inflationary reheating. Specifically, we study inflaton scattering with its decay product, where the couplings involved in this $2 \to 2$ scattering are the same as those in the $1 \to 3$ graviton Bremsstrahlung process. We compute the graviton production rate via such $2 \to 2$ scattering. Additionally, we compare the resulting GW spectrum with that from Bremsstrahlung as well as that from pure $2 \to 2$ inflaton scatterings. For completeness, the GW spectrum from graviton pair production through one-loop induced $1 \to 2$ inflaton decay is also analyzed. With a systematic comparison among the four sources of GWs, we find that $2 \to 2$ inflaton scattering with its decay product can dominate over Bremsstrahlung if the reheating temperature is larger than the inflaton mass. Pure inflaton $2 \to 2$ scattering is typically subdominant compared to Bremsstrahlung except in the high-frequency tail. The contribution from one-loop induced $1 \to 2$ inflaton decay is shown to be suppressed compared to Bremsstrahlung and pure inflaton $2 \to 2$ scattering.
}

\begin{document} 
	\begin{flushright}
		MITP-24-058\\
		December 2024 
	\end{flushright}
	\maketitle
	\flushbottom
	
	\section{Introduction}
	Ultra-high frequency gravitational waves (GWs) can be generated through several processes in the early universe, such as inflationary vacuum fluctuations and non-perturbative preheating after inflation \cite{Caprini:2018mtu}, graviton Bremsstrahlung \cite{Nakayama:2018ptw, Huang:2019lgd, Barman:2023ymn, Barman:2023rpg, Kanemura:2023pnv, Bernal:2023wus, Hu:2024awd, Barman:2024htg}, inflaton scattering \cite{Ema:2015dka, Ema:2016hlw, Ema:2020ggo, Ema:2021fdz}, and fluctuations in the thermal plasma \cite{Ghiglieri:2015nfa, McDonough:2020tqq, Ghiglieri:2020mhm, Ringwald:2020ist, Klose:2022knn, Klose:2022rxh, Ringwald:2022xif, Ghiglieri:2022rfp, Muia:2023wru, Drewes:2023oxg, Ghiglieri:2024ghm},  topological defects as well as evaporation of primordial black holes \cite{Anantua:2008am,  Gehrman:2022imk,Ireland:2023avg, Gehrman:2023esa, Choi:2024acs}. See Ref.~\cite{Aggarwal:2020olq} for a recent review on ultra-high frequency GWs. 
	
	In this work, we revisit the generation of GWs from gravitons production during reheating. Different from existing analyses in the literature \cite{Nakayama:2018ptw, Huang:2019lgd, Barman:2023ymn, Barman:2023rpg, Kanemura:2023pnv, Bernal:2023wus} that consider GWs from gravitons sourced from $1 \to 3$ Bremsstrahlung, we investigate the GW signatures generated from $2 \to 2$ inflaton scattering with its decay product.\footnote{In Ref.~\cite{Klose:2022knn}, a similar process is considered with a non-Abelian gauge field in the final states.} Note that the couplings involved in such $2 \to 2$ scatterings are the same as those in $1 \to 3$ Bremsstrahlung. However, the resulting GW spectra are expected to be different. Firstly, the kinematics of the gravitons produced from $2 \to 2$ scatterings differ from those in $1 \to 3$ Bremsstrahlung. For $1 \to 3$ Bremsstrahlung, the maximum energy of the graviton at emission is half of the inflaton mass. However, for $2 \to 2$ inflaton scatterings with its decay product, the energy of the produced graviton can be equal to the inflaton mass. Consequently, the frequency of  GW spectrum that can be reached from $2 \to 2$ scatterings is expected to be higher than that from $1 \to 3$ processes.  Note that it is possible that  the inflaton decay products thermalize rapidly during reheating if the decay of inflaton is not Planck suppressed \cite{Harigaya:2013vwa}.  Consequently, it is conceivable that $2 \to 2$  scatterings between inflaton and the thermalized decay product  can dominate over $1 \to 3$ Bremsstrahlung if the temperature during reheating is higher than the inflaton mass. Note that the graviton production rate for $2 \to 2$ scatterings is temperature-dependent, as one of the initial states is in thermal equilibrium. In contrast, $1 \to 3$ Bremsstrahlung is a purely non-thermal process with a different graviton production rate. Therefore, depending on the underlying reheating process, the shape of GW spectrum from $2 \to 2$ scatterings is expected to differ from that of $1 \to 3$ Bremsstrahlung.
	
	The goal of this work is to compute the graviton production rate for $2 \to 2$ inflaton scattering with its decay product, obtain the resulting GW spectrum, and further compare it with previous results from $1 \to 3$ Bremsstrahlung \cite{Nakayama:2018ptw, Huang:2019lgd, Barman:2023ymn, Barman:2023rpg, Bernal:2023wus}. This is one objective of this work. Note that for both $1 \to 3$ Bremsstrahlung and $2 \to 2$ inflaton scattering with decay product, only one graviton is produced in the final state. It has been shown that GWs with double gravitons can be sourced from inflaton-inflaton scatterings \cite{Ema:2015dka, Ema:2016hlw, Ema:2020ggo, Ema:2021fdz}. For completeness, we also present the unavoidable GW spectrum from double gravitons via one-loop induced inflaton $1\to 2$ decays during reheating\footnote{We work in the minimally coupled gravity framework, where there is no vertex giving rise to double graviton production at the tree level from inflaton decay. In a modified gravity framework, such as $f(R)$ gravity, a direct vertex between the inflaton and gravitons is possible, allowing gravitons to be generated directly from inflaton decays at tree level \cite{Ema:2021fdz, Tokareva:2023mrt, Koshelev:2022wqj}.}. Finally, we offer a comprehensive comparison of the GW spectra from the aforementioned four sources, which is missing in the literature. Our goal is to identify the dominant process and the corresponding conditions under which it occurs.
	
	The paper is organized as follows. In Sec.~\ref{sec:setup}, we present the model framework. We revisit reheating by considering inflaton bosonic or fermionic decay in Sec.~\ref{sec:reheating}. In Sec.~\ref{sec:rate}, we present the graviton production rates. The energy spectrum of gravitons from different sources is computed in Sec.~\ref{sec:energy_spectrum}. The GW spectrum and a systematic comparison among different sources are provided in Sec.~\ref{sec:GW}. Finally, we summarize the findings of this work in Sec.~\ref{sec:conclusion}.

	\section{The Setup}\label{sec:setup}
	We work in the minimal coupled gravity framework. The is action is give by 
	\begin{align}
		S \supset \int d^4 x\sqrt{-g} \left[\mathcal{L}_{\text{EH}} + \mathcal{L}_\phi+ \mathcal{L}_{\text{int}} + \mathcal{L}_\text{SM} \right]\,,
	\end{align}
	where $\mathcal{L}_{\text{EH}} = \frac{M_P^2}{2}R $ denotes the 
	Einstein-Hilbert term for gravity with $M_P\equiv 1/\sqrt{8\pi G_N}$ being  the reduced Planck mass, $g$ the determinant of the metric $g_{\mu \nu}$, and $R$ the Ricci scalar. The second term
	corresponds the Lagrangian  density for inflaton field $\phi$, given by $ \mathcal{L}_\phi =\left(\frac{1}{2}g^{\mu \nu} \partial_{\mu}\phi \partial_{\nu} \phi -V(\phi)\right) $.  The standard model Lagrangian is denoted as $\mathcal{L}_\text{SM} $, and $\mathcal{L}_{\text{int}}$ incorporates the interactions describing  the inflaton energy transfer to SM degrees of freedom for reheating, which will be specified in the following section. 
	
	In order to obtain the gravitational interaction vertex, we expand the metric $g_{\mu \nu}$ around the Minkowski metric $\eta_{\mu \nu}=(+,-,-,-)$ as \cite{Choi:1994ax}
	\begin{align}\label{eq:expansion}
		g_{\mu \nu} =\eta_{\mu \nu} + \kappa\, h_{\mu \nu} \,,
	\end{align}
	where $\kappa= \frac{2}{M_P}$, and $h_{\mu \nu}$ denotes the spin-2 graviton field with mass dimension one.  From Eq.~\eqref{eq:expansion}, it follows that $\sqrt{-g} \simeq 1+\frac{\kappa}{2} h$ with $ h^{\mu}_{\mu} $ being the trace of the graviton field. With $g_{\mu \nu}$, one can further compute the contravariant form $g^{\mu \nu}$, which appears in the Lagrangian densities. By plugging the expansion back into the action, we obtain the effective couplings between the graviton and the energy-momentum tensor \cite{Choi:1994ax}:
	\begin{align}\label{eq:effective}
		\sqrt{-g} \mathcal{L} \supset \frac{1}{M_P} h_{\mu \nu} \sum_k T_k^{\mu \nu}\,,
	\end{align}
	where $T_k^{\mu \nu}$ denotes the energy-momentum tensor for a particle species $k$, which can be the inflaton, its decay products, as well as Standard Model (SM) degrees of freedom. We note that in order to source gravitons, it is required to have an  anisotropic\footnote{The energy momentum for inflaton reads $T_\phi^{\mu \nu} = \left[\partial^{\mu}\phi \partial^{\nu}\phi -\frac{1}{2} \eta^{\mu \nu} (\partial_\alpha \phi \partial^{\alpha} \phi - m_\phi^2 \phi^2)\right]$, which implies that $T_\phi^{i j} =0$ since inflaton condensate is homogeneous with vanishing anisotropic densities. This also implies  that the amplitude for the Feynman diagram with a graviton-inflaton vertex would vanish.} energy momentum tensor $T^{i j} \neq 0$ with $i,j =1, 2, 3$.
	
	\section{Reheating}\label{sec:reheating}
	After inflation ends, the inflaton rolls down to the minimum of the inflaton potential and starts oscillating there, producing lighter degrees of freedom in the standard model (SM). These daughter particles interact with each other, reaching equilibrium and forming a SM thermal bath \cite{Allahverdi:2010xz, Lozanov:2019jxc}. Thermalization of SM particles proceeds through gauge interactions, while the inflaton is a weakly coupled sector. Consequently, the timescale for thermalization is expected to be much shorter than the lifetime of the inflaton (or the duration of reheating phase), implying a rapid thermalization \cite{Chung:1998rq, Kolb:2003ke}.  The detailed processes\footnote{They include energy dissipation of  high-energy inflaton decay products from cascaded splittings, which are dominated by scatterings with small angles and small momentum transfers due to collinear effects. This  leads to a thermalization rate of the form $\Gamma_{\text{th}} \sim \alpha^2 T \sqrt{T/m_\phi}$, with $\alpha = g^2/(4\pi)$ denoting the gauge coupling strength~\cite{Harigaya:2013vwa, Mukaida:2022bbo}.} of thermalization of inflaton decay products during reheating have been thoroughly investigated in the literature \cite{Allahverdi:2000ss, Davidson:2000er, Kurkela:2011ti, Harigaya:2013vwa, Garcia:2018wtq, Drees:2021lbm, Drees:2022vvn, Mukaida:2022bbo, Chowdhury:2023jft}. 
	For typical gauge couplings, the thermalization condition of the SM plasma, $\Gamma_{\text{th}} > H$ (with $H$ denoting the Hubble parameter  and $\Gamma_{\text{th}}$ the scattering rate during reheating), is quickly satisfied well before the end of reheating and almost instantaneously~\cite{Harigaya:2013vwa}.

	Throughout this work, we assume the inflaton oscillates around a quadratic potential\footnote{For an inflaton potential steeper than quadratic, the inflaton features a self-interaction, which could lead to inflaton fragmentation and breakdown of coherent oscillations. See Refs.~\cite{Lozanov:2016hid, Lozanov:2017hjm, Garcia:2023eol, Garcia:2023dyf, Garcia:2024zir} for recent studies in this direction.} of the form:
	\begin{align}
		V(\phi) = \frac{1}{2} m_\phi^2 \phi^2\,,
	\end{align}
	where $m_\phi$ denotes the inflaton mass.
	We note that a quadratic potential during reheating could originate from viable inflation models constrained by current Cosmic Microwave Background experiments, such as Starobinsky inflation \cite{Starobinsky:1980te}, the $\alpha$-attractor model \cite{Kallosh:2013hoa, Kallosh:2013maa}, and the simple and well-motivated polynomial inflation \cite{Drees:2021wgd, Drees:2022aea}.
	
	We consider a scenario where the inflaton decays into a pair of lighter bosons (e.g., the Higgs field in the Standard Model) or vector-like fermions through the following trilinear couplings:
	\begin{align}\label{eq:int1}
		\mathcal{L}_\text{int} \supset \mu\, \phi\, |\varphi|^2 + y\, \phi  \bar{\psi}\, \psi\,,
	\end{align}
	with which the decay rates read
	\begin{align}\label{eq:phiFF}
		\Gamma_\phi\simeq 
		\begin{cases}
			\frac{m_\phi}{8\,\pi} \left(\frac{\mu}{m_\phi}\right)^2 &\text{bosonic decay}\,,
			\\
			\frac{ m_\phi}{8\pi}y^2& \text{fermionic decay}\,,
		\end{cases}
	\end{align}
	where we have neglected masses for the daughter particles. Note that the couplings between the inflaton and the daughter fields generate effective mass terms ($\propto \phi$) for the daughter fields, modifying the decay kinematics. To account for this effect, it is necessary to average over inflaton oscillations, which results in effective couplings, $\mu_{\text{eff}}$ and $y_{\text{eff}}$, for bosonic and fermionic decays, respectively. For a quadratic inflaton potential, it has been shown that this effect is moderate, with $\mu_{\text{eff}} \simeq \mu$ and $y_{\text{eff}} \simeq y$ \cite{Ichikawa:2008ne, Garcia:2020wiy}. With the decay rate, one can track the evolution of energy densities, namely the inflaton energy density $\rho_\phi \equiv \frac{\dot{\phi}^2}{2} + V(\phi)$ and the radiation energy density $\rho_R \equiv \frac{g_*\, \pi^2}{30} T^4$ with $g_*$ being the degrees of freedom in the thermal bath. The evolution equations are given by the following Boltzmann equations:
	\begin{align}
		&\frac{d\rp}{dt} + 3 H\,\rp = -\Gamma_\phi\, \rp\,, \label{eq:rhophi}\\
		&\frac{d\rR}{dt} + 4 H\, \rR = + \Gamma_\phi \, \rp \label{eq:rhoR},
	\end{align}
	where the Hubble parameter $H$ is defined via $H \equiv \dot{a}/a$ with $a$ being the scale factor. Using the Friedmann equation, it follows that
	\begin{align}\label{eq:Hubble1}
		H^2 = \frac{\rR + \rp}{3 M_P^2}\,.
	\end{align}
	The solution for Eq.~\eqref{eq:rhophi} with scale factor $a$ as variable  is given by 
	\begin{align}\label{eq:rhophi_sol} 
		\rp(a) & \simeq \rp(\aend)\left( \frac{\aend}{a}\right)^3 \text{e}^{-\frac{2\Gamma_\phi}{3 H} \left[1 - \left(\frac{\aend}{a}\right)^{3/2}\right]}\nonumber \\
		&\simeq  3 H^2_{\text{inf}} M_P^2\left( \frac{\aend}{a}\right)^3\,,    
	\end{align} 
	where in the last line $\Gamma_\phi \ll H$ is assumed. Here, $\aend$ denotes the scale factor at the end of inflation, and $\arh$ corresponds to the scale factor at the end of reheating. Note that the inflaton lifetime is $t_\phi \simeq \frac{1}{\Gamma_\phi}$ and Hubble parameter scales as $H(t)\simeq \frac{2}{3t}$ during during reheating, leading $H(\Trh) =\frac{2}{3} \Gamma_\phi$ at the end of reheating. This gives
	\begin{align}\label{trh}
		\Trh = \sqrt { \frac {2} {\pi} } \left( \frac {10} {\gs} \right)^{1/4}
		\sqrt {M_P\, \Gamma_\phi}\,.
	\end{align}
	With Eq.~\eqref{eq:rhophi_sol} we can further solve Eq.~\eqref{eq:rhoR} and obtain the solution for $\rR$:
	\begin{align}\label{eq:rho_R_sol}
		\rR(a)=\frac{6}{5} M_P^2 \, \Gamma_\phi \, \Hinf\left(\frac{\aend}{a}\right)^{\frac{3}{2}}\left[1-\left(\frac{\aend}{a}\right)^{5 / 2}\right]\,.
	\end{align}
	The inflationary scale $\Hinf$ is a inflation model dependent parameter,\footnote{The recent BICEP/KecK 2018 constraints on the tensor-to-scalar ratio $r<0.035$  \cite{BICEP:2021xfz} implies that  $\Hinf < 2\cdot10^{-5}\, M_P$.} and can be written as 
	\begin{align}\label{eq:Hinf}
		\frac{H_{\text {inf}}}{M_P}=1.0\cdot10^{-5}\left(\frac{\mathcal{P}}{2.1\cdot10^{-9}}\right)^{1 / 2}\left(\frac{r}{0.01}\right)^{1 / 2}\,,
	\end{align}
	where $r$ denotes the tensor-to-scalar ratio and  $\mathcal{P}$  the scalar power spectrum. In this work,  we  remain agnostic for inflation models. We will consider  the central value for the power spectrum $\mathcal{P} =2.1\cdot10^{-9}$  from Planck 2018 \cite{Planck:2018vyg} and take the recent BICEP/Keck 2018 constraint on the tensor-to-scalar ratio   $r<0.035$ \cite{BICEP:2021xfz} into account.
	
	From Eq.~\eqref{eq:rho_R_sol} we find $\rR(a)$  maximizes at $a =\amax =(8/3)^{2/5} \aend$,  corresponding to a maximum temperature:
	\begin{align}\label{eq:Tmax}
		\Tmax^4=\frac{60}{\pi^2 g_{\star}}\left(\frac{3}{8}\right)^{8 / 5} M_P^2\, \Gamma_\phi\, \Hinf\,.
	\end{align}
	Using $H(\Trh) =\frac{2}{3} \Gamma_\phi$, we can further rewrite Eq.~\eqref{eq:Tmax} as
	\begin{align}\label{eq:Tmax1}
		\frac{\Tmax}{\Trh}=\left(\frac{3}{8}\right)^{2 / 5}\left[\frac{H_{\text{inf}}}{H(\Trh)}\right]^{1 / 4}\,,
	\end{align}
	with which one can further write $\Tmax$ as function of $r$ and $\Trh$
	\begin{align}\label{eq:Tmax2}
		\Tmax = 1.4 \cdot10^{15} \left(\frac{r}{0.01}\right)^{1/8} \left(\frac{\Trh}{10^{13}~\text{GeV}}\right)^{1/2} ~\text{GeV}\,,
	\end{align}
	after using Eq.~\eqref{eq:Hinf}. It is clear that $\Tmax$  can be (much) larger than $\Trh$ \cite{Giudice:2000ex}. The expression for $\Tmax$ presented in Eq.~\eqref{eq:Tmax2} would be useful when we compute the GW amplitude, where the dilution effects depend on $\Tmax$ as will be explained later.
	
	Before closing this section, we note that non-perturbative preheating is subdominant to perturbative decay in our framework. Note that the trilinear coupling $\mu\, \phi\, |\varphi|^2$ can lead to a tachyonic squared-mass $m^2_{\varphi} \sim \mu\, \phi$ for the daughter field $\varphi$ once the inflaton crosses zero and becomes negative during oscillations, tending to make preheating efficient. However, $\varphi$ in our setup is the Higgs field with a self-interaction $\lambda_\varphi\, \varphi^4$, which gives rise to a positive squared mass $m^2_{\varphi} \sim \lambda_{\varphi} \left\langle \varphi^2 \right \rangle$ with $\left\langle \varphi^2 \right \rangle$ being the variance of the produced Higgs mode. Such back-reaction counteracts the tachyonic instability, making preheating inefficient \cite{Dufaux:2006ee}. On the other hand, for inflaton decays to fermions, the Pauli blocking effects forbid preheating from being efficient \cite{Peloso:2000hy}.   Finally, we note that the purely gravitational effect is negligible for reheating in the present setup. Within the minimally coupled gravity framework, it has been shown that for this mechanism to be efficient, the inflaton potential must be relatively steep, specifically following $V(\phi) \sim \phi^p$ with $p > 9$ during the reheating phase~\cite{Clery:2021bwz, Haque:2022kez, Barman:2022qgt, Haque:2023yra}.
	\section{Graviton Production during Reheating}\label{sec:rate}
	In this section, we present the graviton production rates for $1\to 3$ Bremsstrahlung, $2\to 2$ scatterings and one-loop inflaton decays.
	\subsection{\texorpdfstring{$1\to 3$}{1 to 3} Bremsstrahlung}
	\begin{figure}[ht!]
		\def\sepf{0.15}
		\centering
		\includegraphics[scale=\sepf]{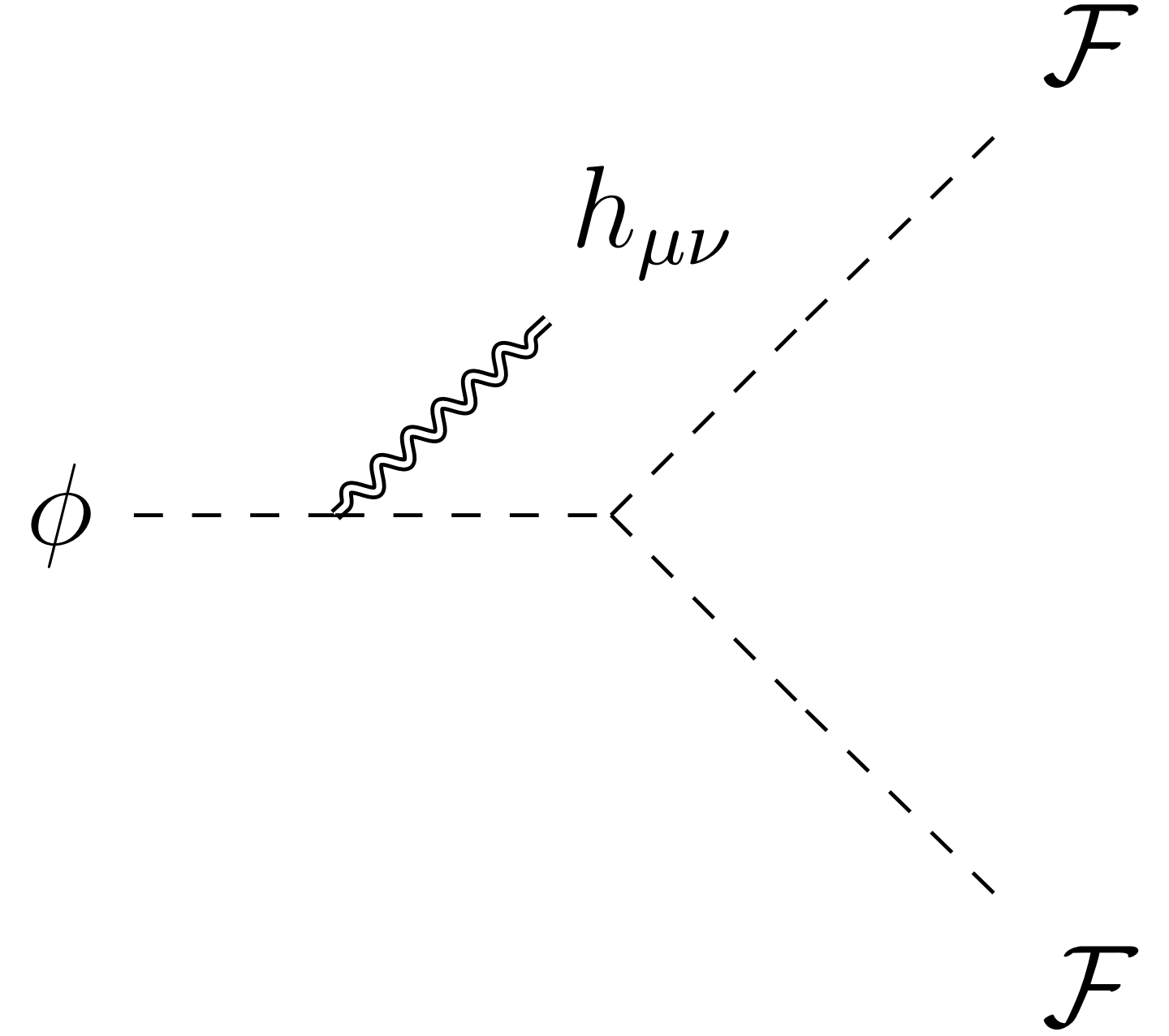}
		\includegraphics[scale=\sepf]{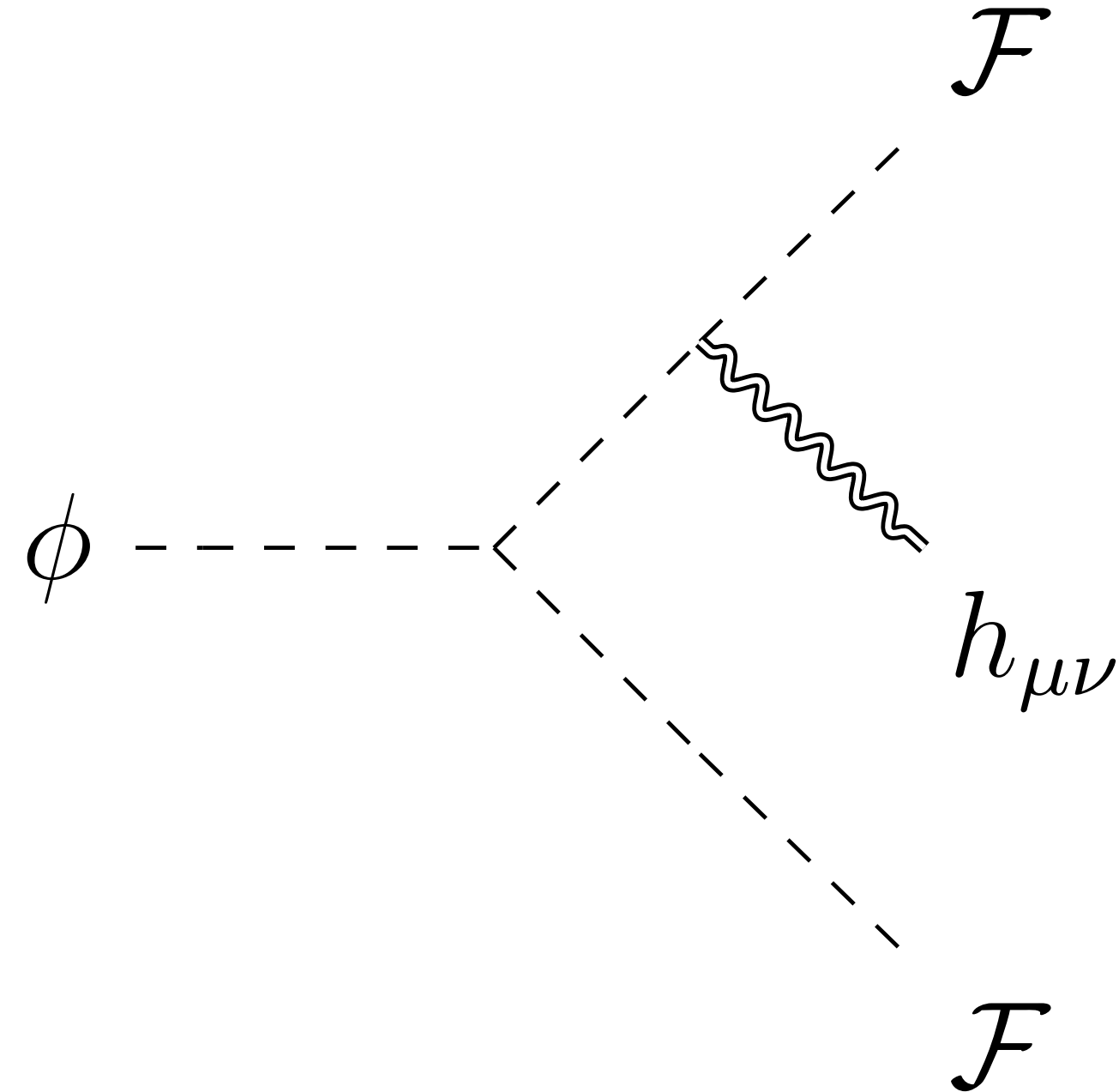}
		\includegraphics[scale=\sepf]{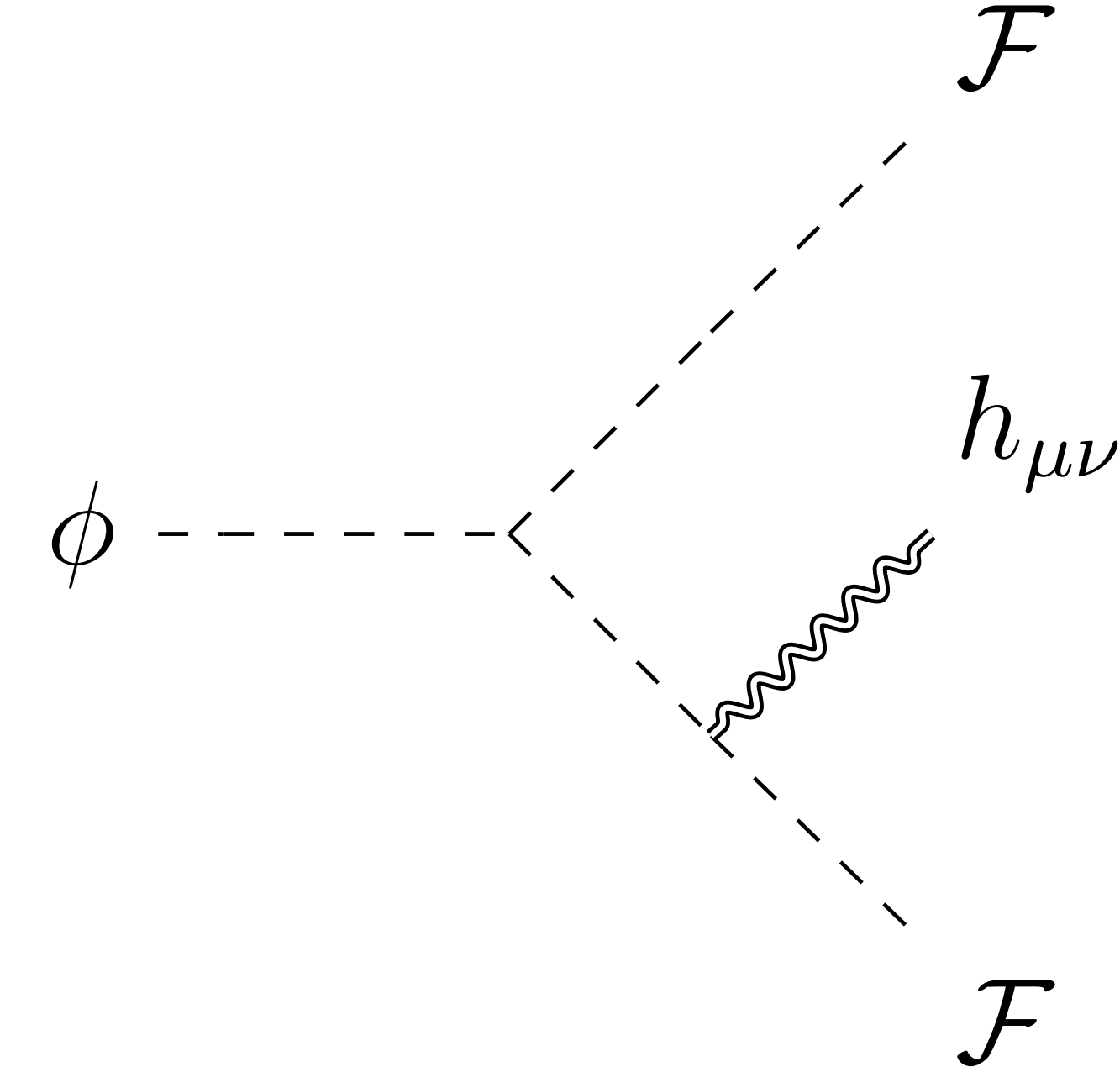}
		\includegraphics[scale=\sepf]{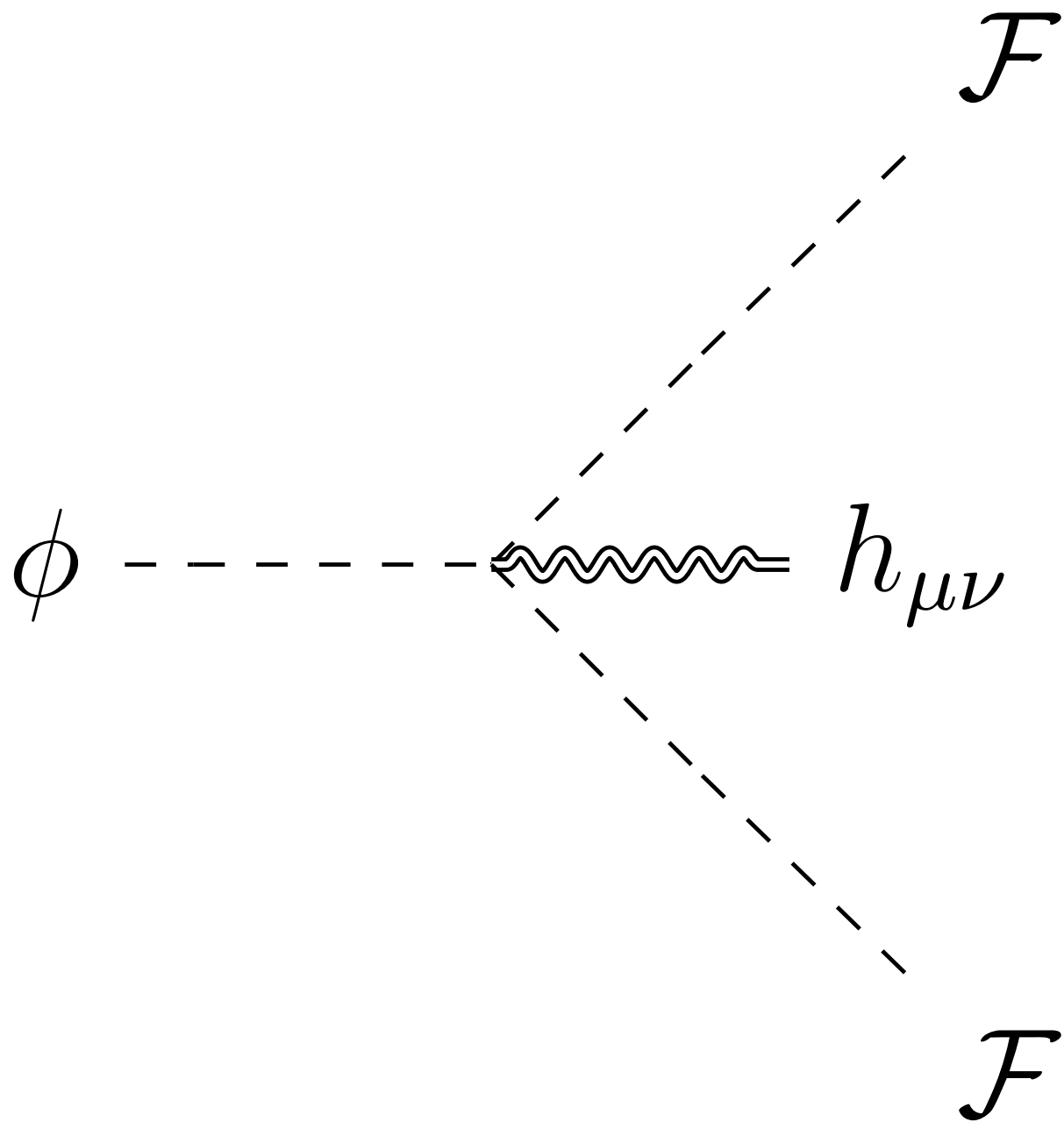}
		\caption{Gravitons production via $1\to 3$ decay with the final states $\mathcal{F} = \{\psi, \varphi\}$.}
		\label{fig:diagram}
	\end{figure} 
	Due to the effective coupling Eq.~\eqref{eq:effective}, gravitons can be produced during reheating via a Bremsstrahlung process as shown by the Feynman diagrams in Fig.~\ref{fig:diagram}. The first diagram arises from the coupling $T_\phi^{\mu \nu} h_{\mu \nu}$, and the corresponding amplitude vanishes since the energy momentum tensor for inflaton condensate is not anisotropic. The scattering amplitude for last diagram vanishes as it contains a term $\eta_{\mu \nu} \epsilon^{\mu \nu} =0$ with 
	$\epsilon^{\mu \nu}$ denoting the graviton polarization tensor. By summing over the contributions from the second and third diagrams, the production rates of graviton with an energy $\Eom$ (or parameter $x=\Eom/m_\phi$) are shown to be \cite{Barman:2023ymn}
	\begin{align}\label{eq:differential_rate}
		\frac{d\Gamma_g^{1\to 3}}{dE_\omega}\simeq 
		\begin{cases}
			\frac{1}{64\,\pi^3} \left(\frac{\mu}{M_P}\right)^2 \frac{(1 - 2 x)^2}{x} &\text{bosonic decay}\,,
			\\
			\frac{y^2}{64\,\pi^3} \left(\frac{m_\phi}{M_P}\right)^2 \frac{(1 - 2 x)}{x} \left[2 x\, (x - 1)  + 1\right]& \text{fermionic decay}\,,
		\end{cases}
	\end{align}
	We note that a graviton can carry a maximum of half of the inflaton energy, which occurs when the daughter particle mass approaches zero. In such a case, the differential decay rate tends toward zero as the phase space closes. This is why the differential decay rate goes to zero when $x \to 1/2$. Moreover, we note that when $x \to 0$, the spectrum $\frac{d\Gamma_g^{1\to 3} }{dE_\omega}$ diverges, which is a well-known feature of (infrared) graviton Bremsstrahlung \cite{Weinberg:1965nx}. To address such divergence, it is necessary to include vertex and self-energy diagrams\footnote{This is similar to the situation in QED.} \cite{Barker:1969jk}. 
	Here, our main interest lies in quantities proportional to $\frac{d\Gamma_g^{1\to 3} }{dE_\omega} E_\omega$, so that the divergence is not problematic.

	\subsection{\texorpdfstring{$2\to 2$}{2 to 2}  Scattering}
	\subsubsection{Inflaton and Decay Product Scattering}
\begin{figure}[t!]
		\def\sepf{0.15}
		\centering
		\includegraphics[scale=\sepf]{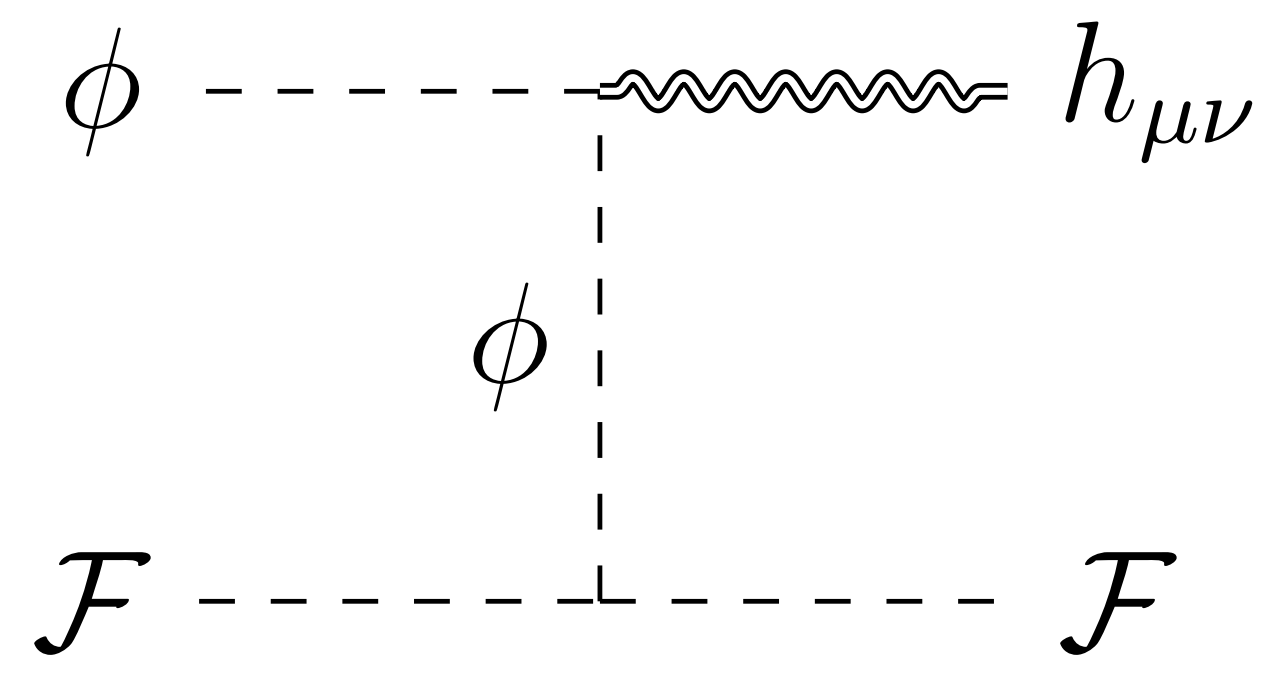}
		\includegraphics[scale=\sepf]{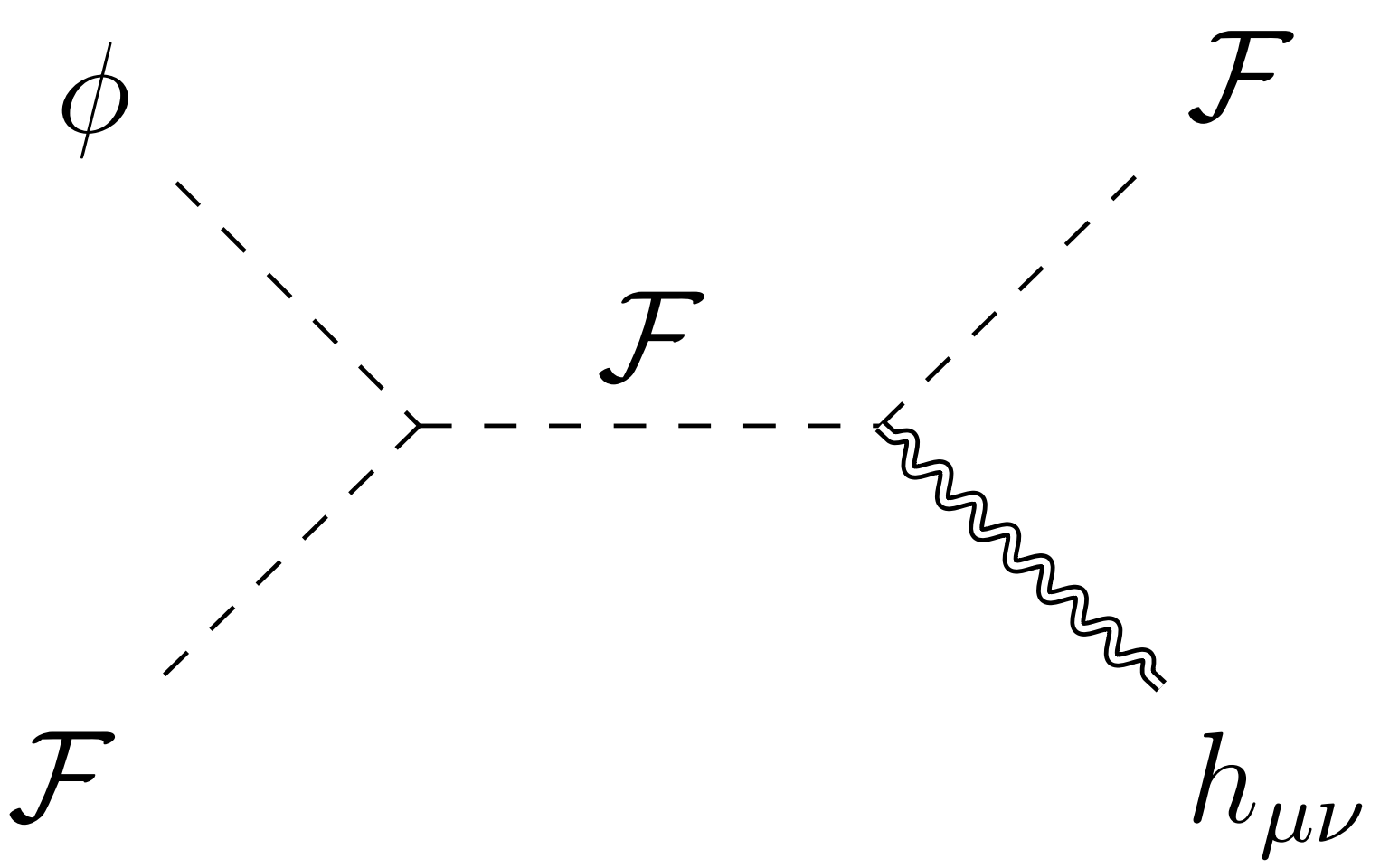}
		\includegraphics[scale=\sepf]{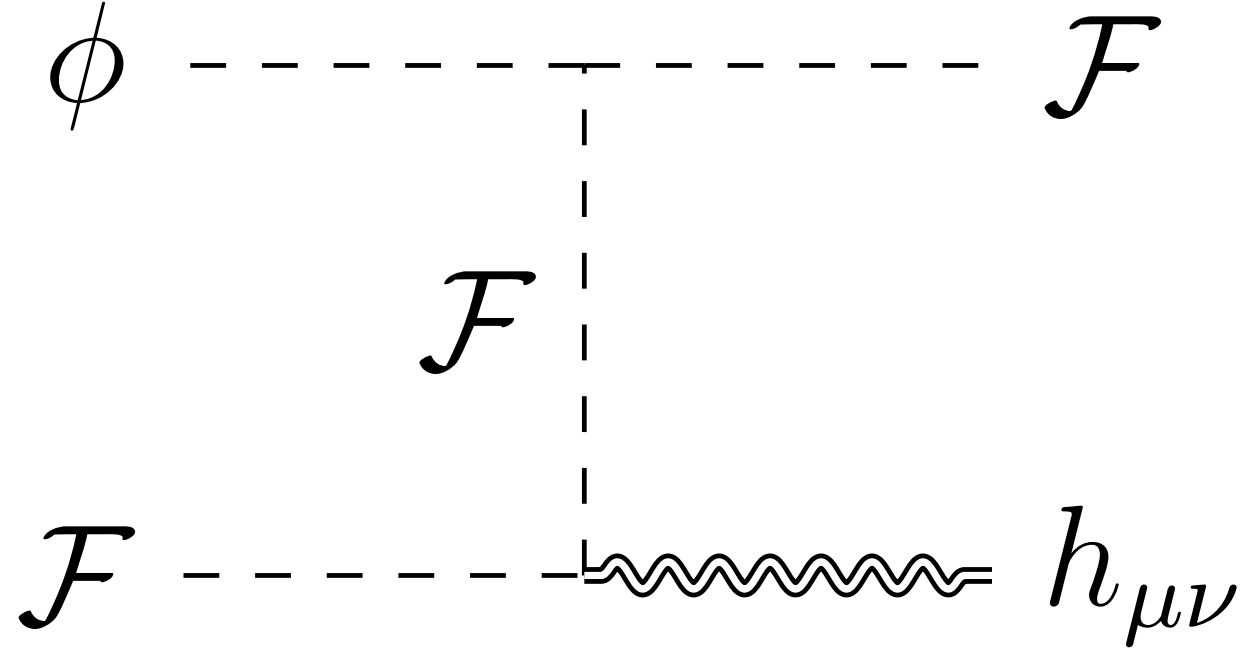}
		\includegraphics[scale=\sepf]{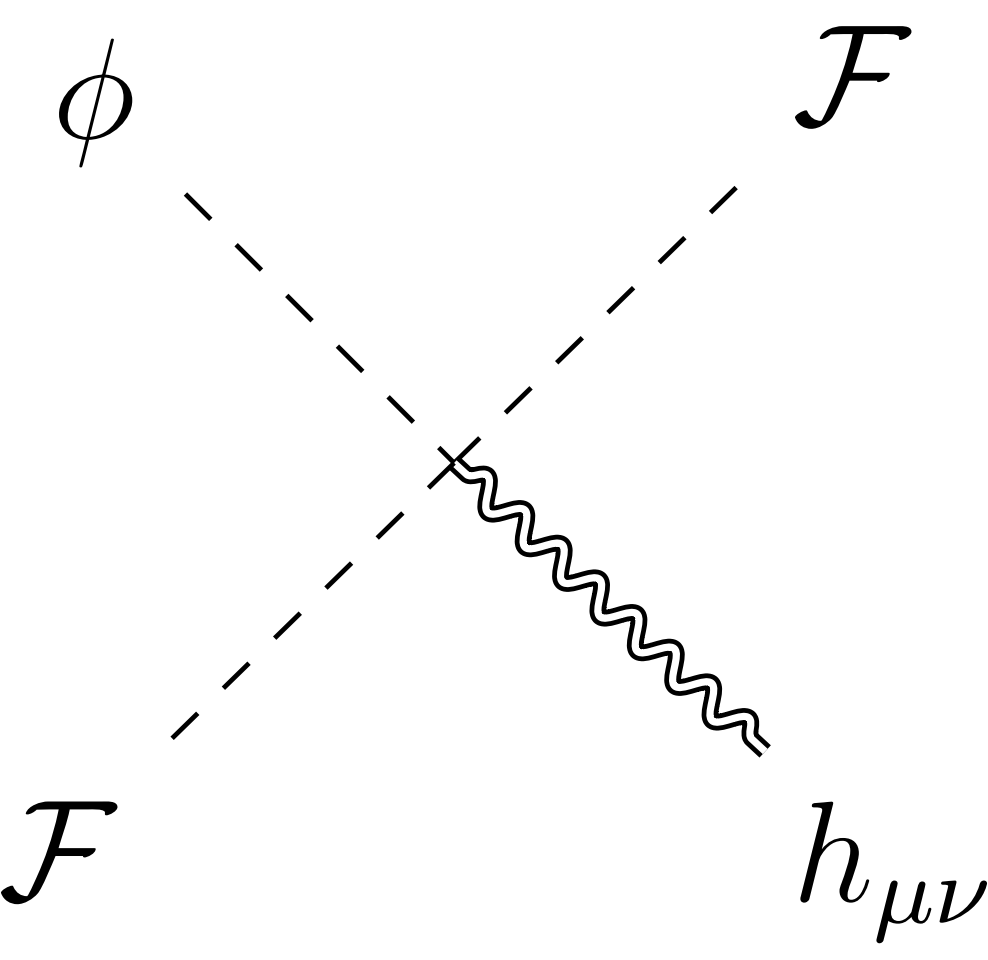}
		\caption{Gravitons production via $2\to 2$ scattering with particles $\mathcal{F} = \{\psi, \varphi\}$ shown in Fig.~\ref{fig:diagram}.}
		\label{fig:2-2diagram}
	\end{figure}

	Along the lines of graviton production from $1 \to 3$ Bremsstrahlung shown in the previous section, we note that during reheating there will be inevitable graviton production via $2 \to 2$ scattering between the inflaton and daughter particles, as shown in Fig.~\ref{fig:2-2diagram}. In this section, different from previous studies \cite{Nakayama:2018ptw, Huang:2019lgd, Barman:2023ymn, Barman:2023rpg, Kanemura:2023pnv, Bernal:2023wus}, we investigate graviton production including inflaton scattering with its decay products. Note that the couplings involved are the same as those for $1 \to 3$ decay, which can be seen from Fig.~\ref{fig:diagram} and Fig.~\ref{fig:2-2diagram}. Here, we aim to compute the graviton production rates from $2 \to 2$ scattering and then compare the corresponding GW spectrum with that from $1 \to 3$ Bremsstrahlung.

	The $2 \to 2$ scattering rate is given by
	\begin{align}\label{eq:2-2_rate_def}
		\Gamma_g^{2 \to 2} = \frac{n_\mathcal{F}}{32\,\pi\,m_\phi\,  E_\mathcal{F}} |\mathcal{M}|^2\,, 
	\end{align} 
	where $n_\mathcal{F}$ denotes the number density of particle $\mathcal{F}$, $E_\mathcal{F}$ the energy, and $\mathcal{M}$ the $2 \to 2$ scattering matrix element for graviton production for the diagram shown in Fig.~\ref{fig:2-2diagram}. The detailed computations are presented in Appendix \ref{sec:appA}. We find that the rates for the $2 \to 2$ scattering between the inflaton and its decay products are given by:
	\begin{align}\label{eq:2-2_rate}
		\Gamma_g^{2 \to 2} \simeq 
		\begin{cases}
			\frac{1}{48 \pi^3 }\frac{\mu^2\, T^2}{M_P^2\,m_\phi} & \text{inflaton-boson}\,, \\
			\frac{3\,y^2}{4 \pi^3 } \frac{T^4}{M_P^2\, m_\phi} & \text{inflaton-fermion}\,,
		\end{cases}
	\end{align}
	where we have included the four degrees of freedom of the $\mathcal{F}$ particle.  The first line of Eq.~\eqref{eq:2-2_rate} corresponds to the case where the inflaton scatters with a thermalized bosonic decay product $\varphi$, and the second line corresponds to the inflaton scattering with a thermalized $\psi$. We note that scattering between the inflaton and non-thermalized $\varphi$ and $\psi$ particles can also occur in the pre-thermalization phase. However, contributions to graviton production from this much earlier phase are highly suppressed due to significant entropy dilution, as we are primarily interested in the quantities at the end of reheating.

	Several comments are in order before closing this subsection.  First, we note that the $2 \to 2$ scattering rate $\Gamma^{2 \to 2}$ in Eq.~\eqref{eq:2-2_rate} can be larger than the $1 \to 3$ rate $\Gamma^{1 \to 3}$ in Eq.~\eqref{eq:differential_rate} if $T \gg m_\phi$. Indeed, we find $\Gamma^{2 \to 2} \propto \left(\frac{T}{m_\phi}\right)^2\Gamma^{1 \to 3}$ for bosonic case, and $\Gamma^{2 \to 2} \propto \left(\frac{T}{m_\phi}\right)^4\Gamma^{1 \to 3}$ for fermionic case. Besides, it is also interesting to note that the graviton production rate $\Gamma^{2 \to 2}$ features time-dependence via temperature evolution during reheating. Due to the different scaling of $\Gamma^{2 \to 2}$ on $T$ and $m_\phi$ in the two cases, it is expected that the graviton production can be more efficient for inflaton scattering with bosonic decay products if\footnote{Note that the couplings in Eq.~\eqref{eq:2-2_rate} can be rewritten with $\Gamma_\phi$ and inflaton mass via Eq.~\eqref{eq:phiFF}. The first line of Eq.~\eqref{eq:2-2_rate}  can be expressed as $\Gamma_g^{2 \to 2} =\frac{\Gamma_\phi}{6\,\pi^2 }\frac{T^2}{M_P^2}$ for bosonic case, while the second line can be rewritten as $\Gamma_g^{2 \to 2} =\frac{6\Gamma_\phi}{\pi^2 }\frac{T^4}{M_P^2m_\phi^2}$ for fermionic case. We remind the reader that for a fixed reheating  temperature, the inflaton decay rate is fixed to be $ \Gamma_\phi = \frac{3}{2}H(\Trh)$.} $\Trh > m_\phi$. This gives rise to possible distinctions on the GW spectrum as will been discussed in next sections.  Finally, depending on when the gravitons are produced during reheating, they receive different redshifts, leading to a graviton energy spectrum at the end of reheating. We will come back to the energy spectrum  of graviton in next section.
	\subsubsection{Inflaton and Inflaton Scattering}
	\begin{figure}[h!]
		\def\sepf{0.15}
		\centering
		\includegraphics[scale=\sepf]{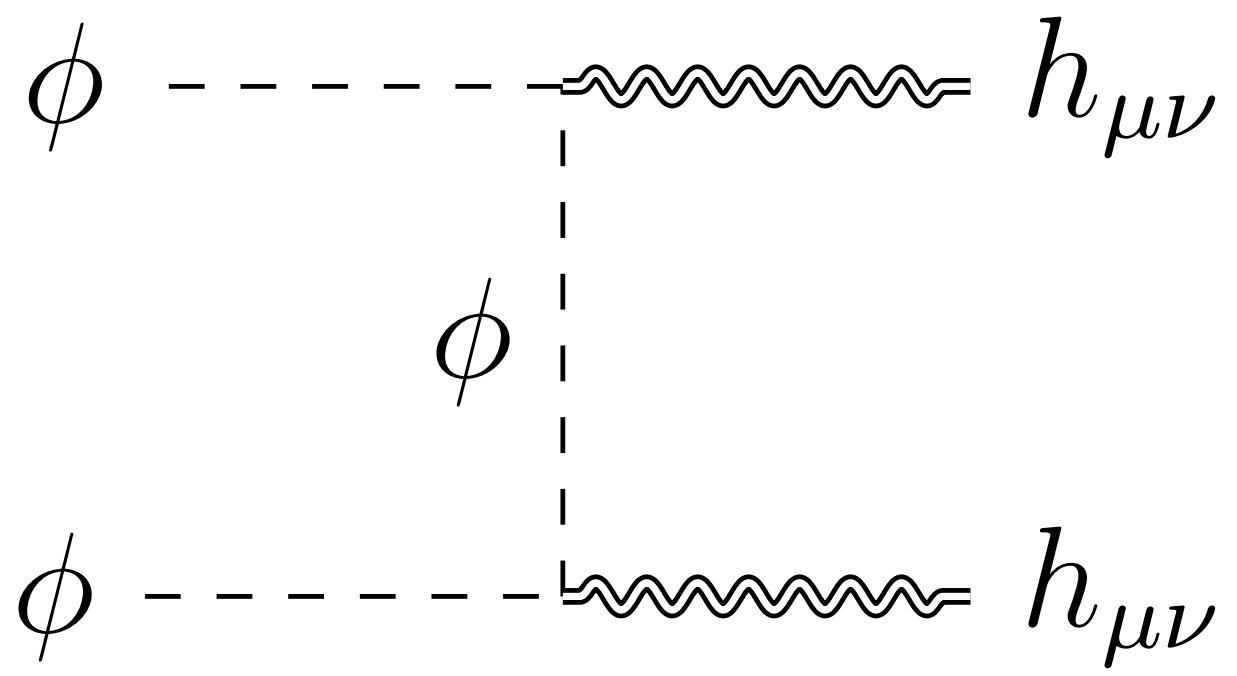}
		\includegraphics[scale=\sepf]{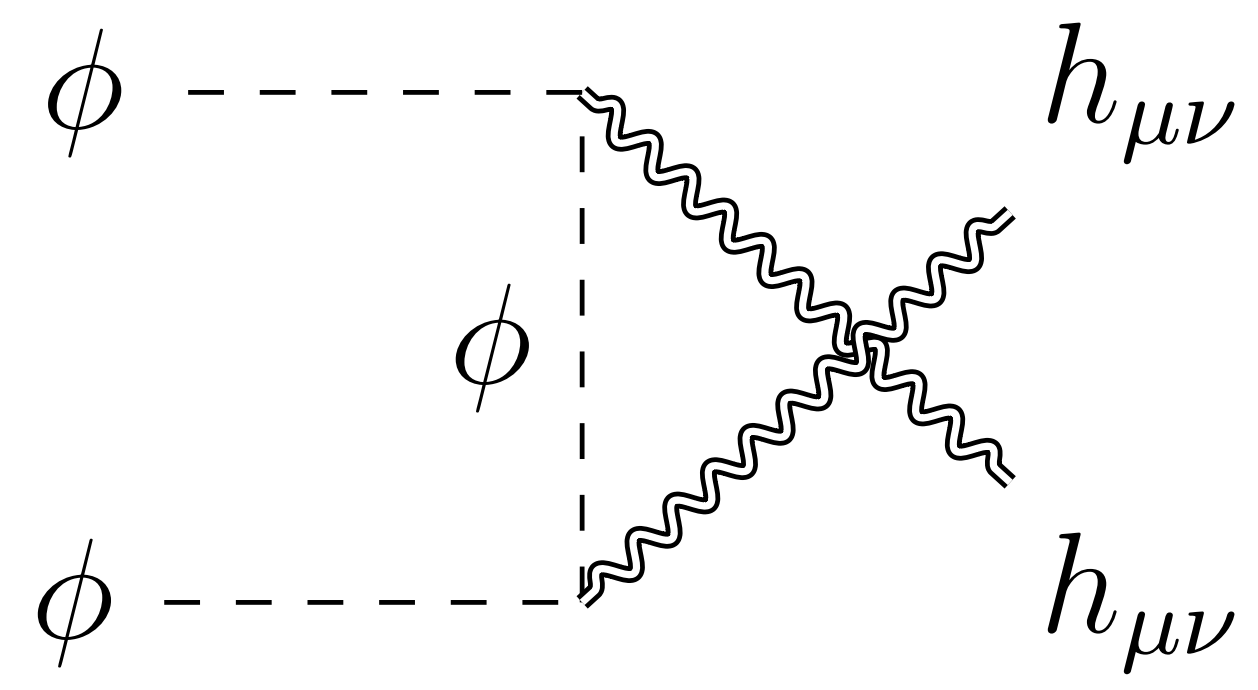}
		\includegraphics[scale=\sepf]{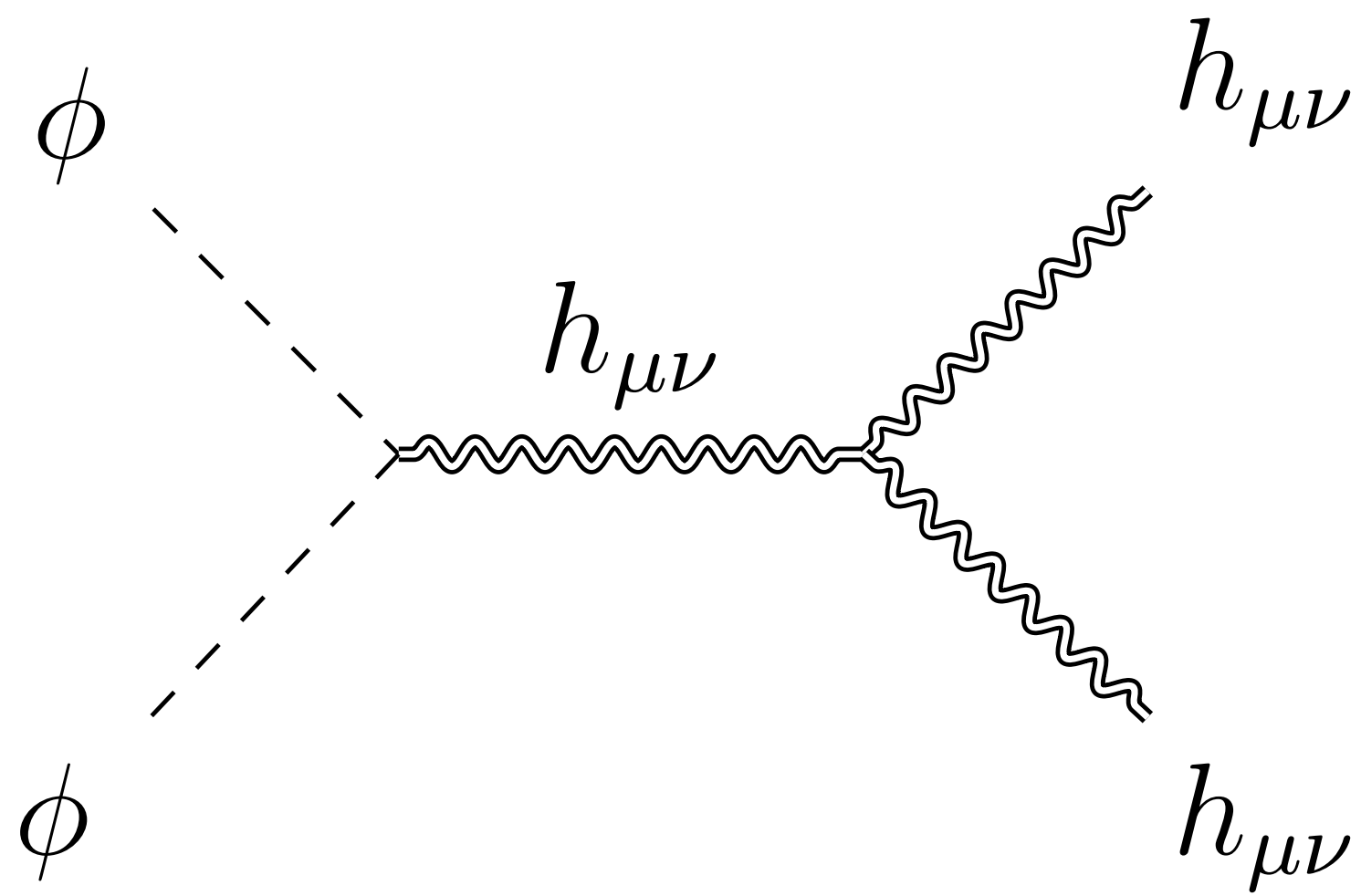}
		\includegraphics[scale=\sepf]{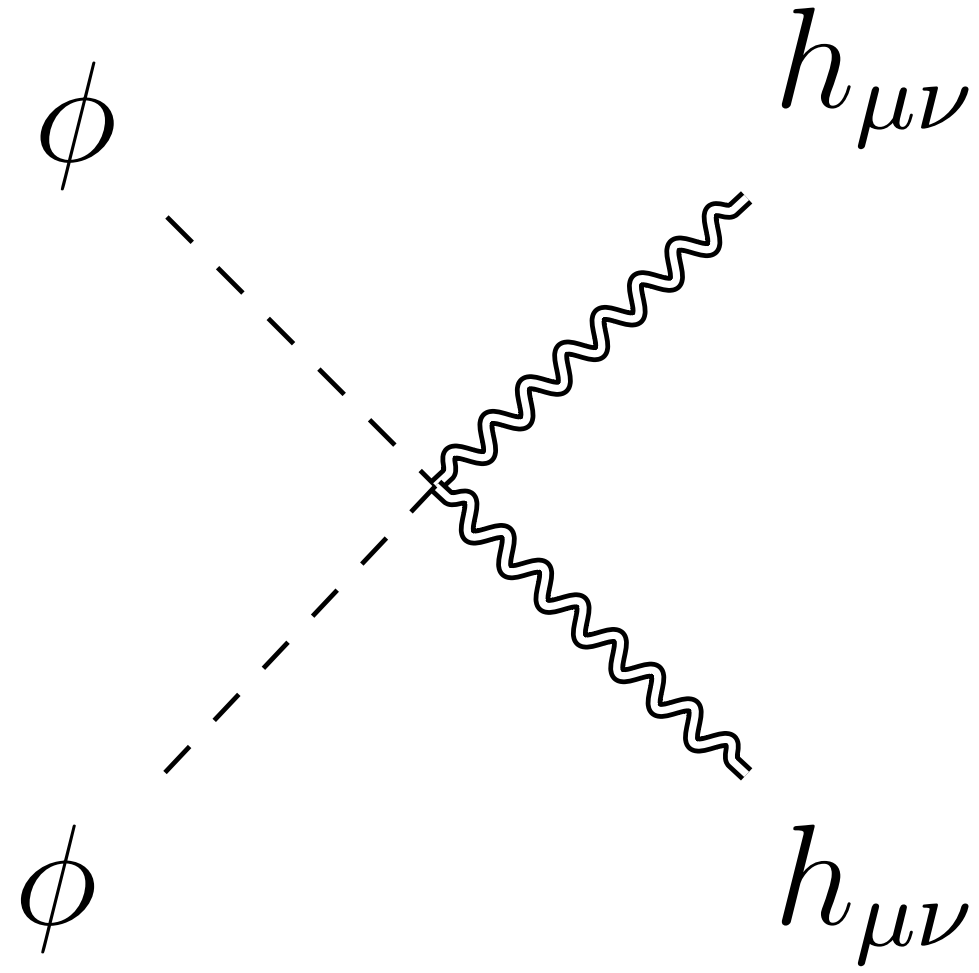}
		\caption {Feynman diagram for pair production of gravitons from inflaton annihilation.}
		\label{fig:gra_pair_ann}
	\end{figure} 
	During reheating, gravitons can also be produced from inflaton-inflaton annihilation \cite{Ema:2015dka, Ema:2016hlw, Ema:2020ggo, Choi:2024ilx}. The diagrams are shown in Fig.~\ref{fig:gra_pair_ann} with double gravitons in the final state. The matrix element for the first two diagrams vanishes as the inflaton behaves as non-relativistic with vanishing three-momentum \cite{Barman:2023ymn}. After summing over the contributions from the third and fourth diagrams, the total graviton production rate is shown to be
	\cite{Choi:2024ilx}:
	\begin{align}\label{eq:2-2_inflaton}
		\Gamma_g^{2\to 2} = \frac{\rp}{m_\phi}\frac{m^2_\phi}{32\pi\, M_P^4} = \frac{\rp\, m_\phi}{32\pi\, M_P^4}\quad \text{inflaton-inflaton}\,,
	\end{align}
	where $\rp$ denotes the inflaton energy density, and $ \frac{\rp}{m_\phi}$ corresponds to the inflaton number density.
	
	\subsection{One-loop Induced \texorpdfstring{$1\to 2$}{1 to 2}  Decay}
\begin{figure}[ht!]
		\def\sepf{0.15}
		\centering
		\includegraphics[scale=\sepf]{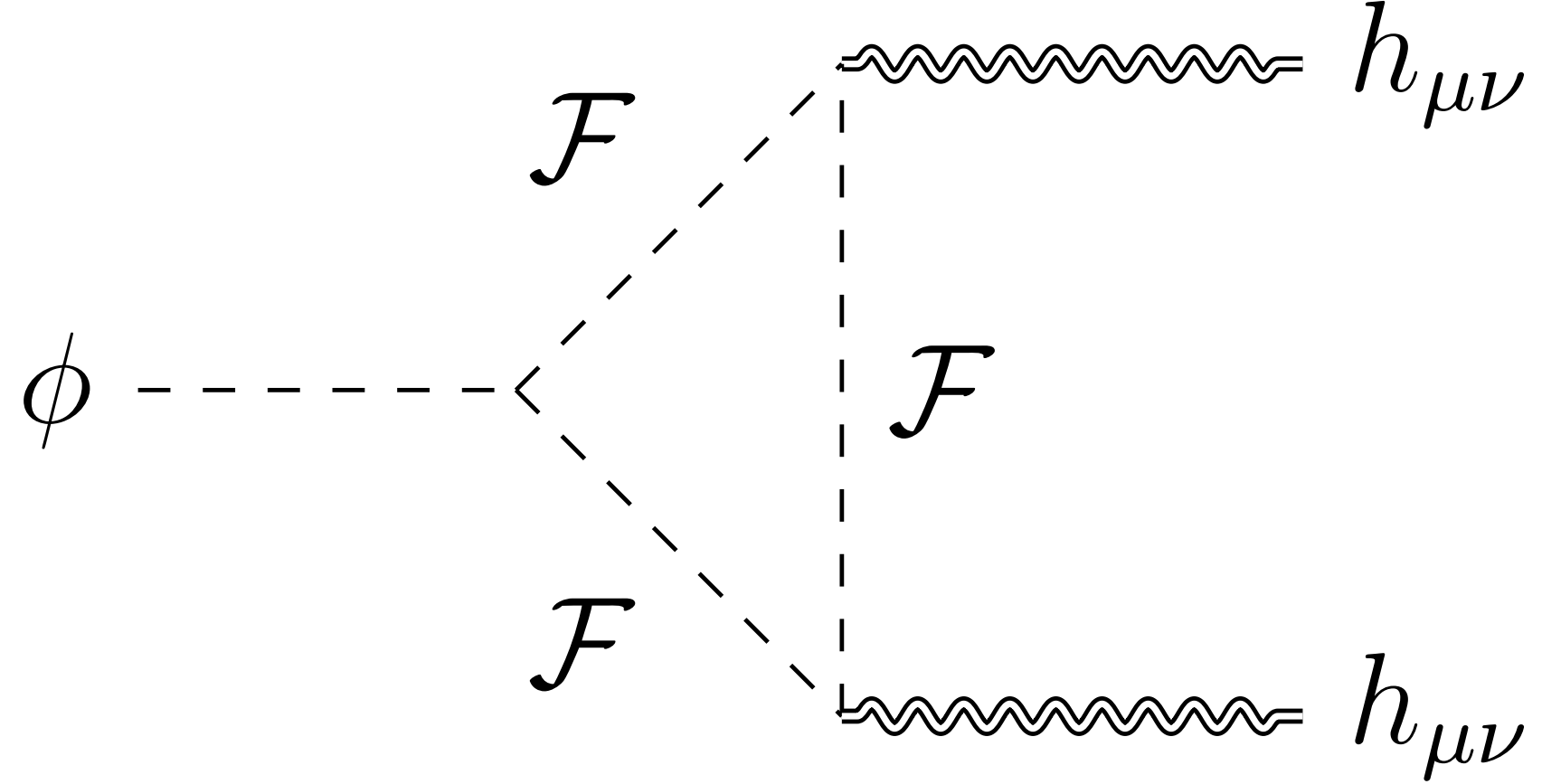}
		\includegraphics[scale=\sepf]{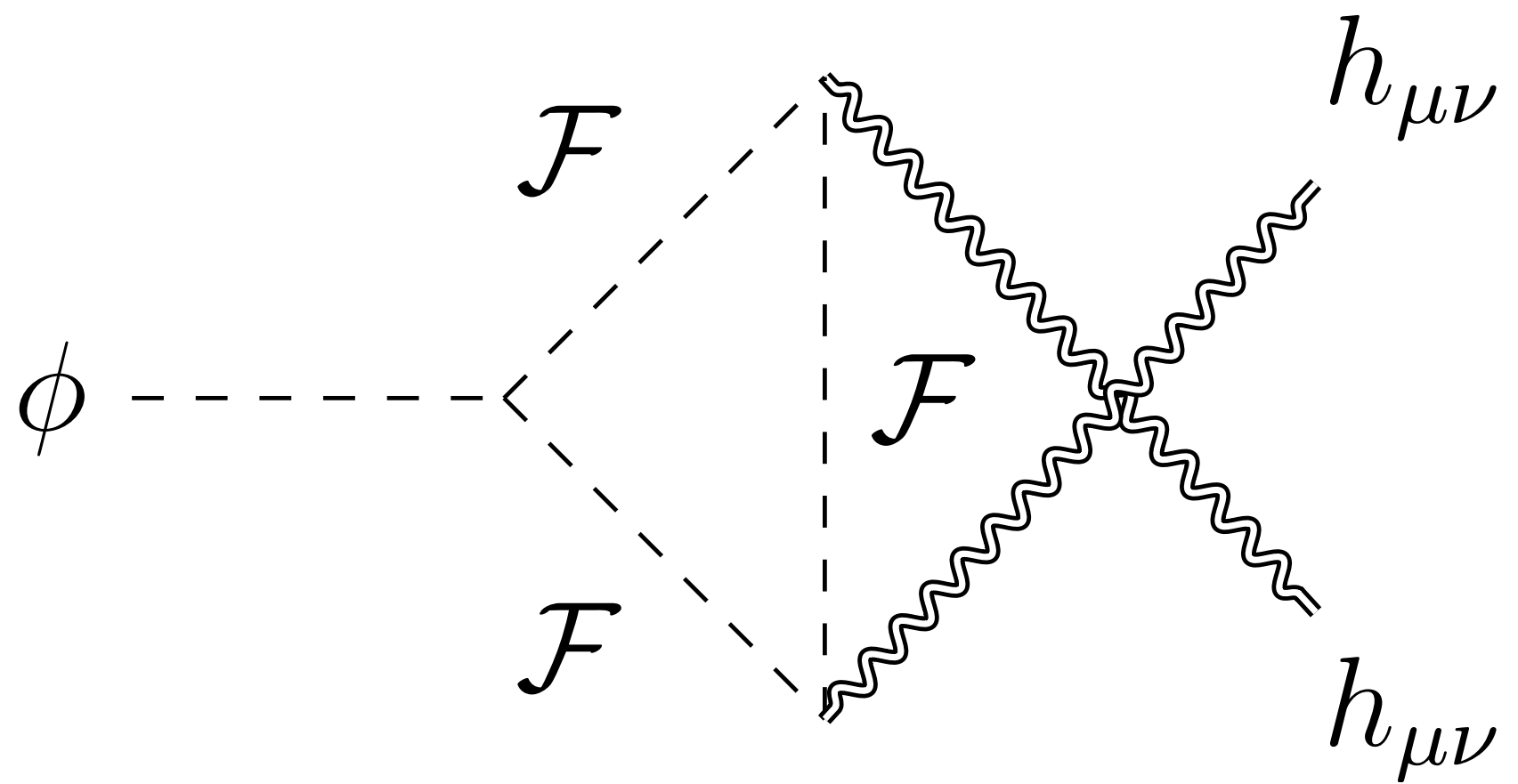}
		\includegraphics[scale=\sepf]{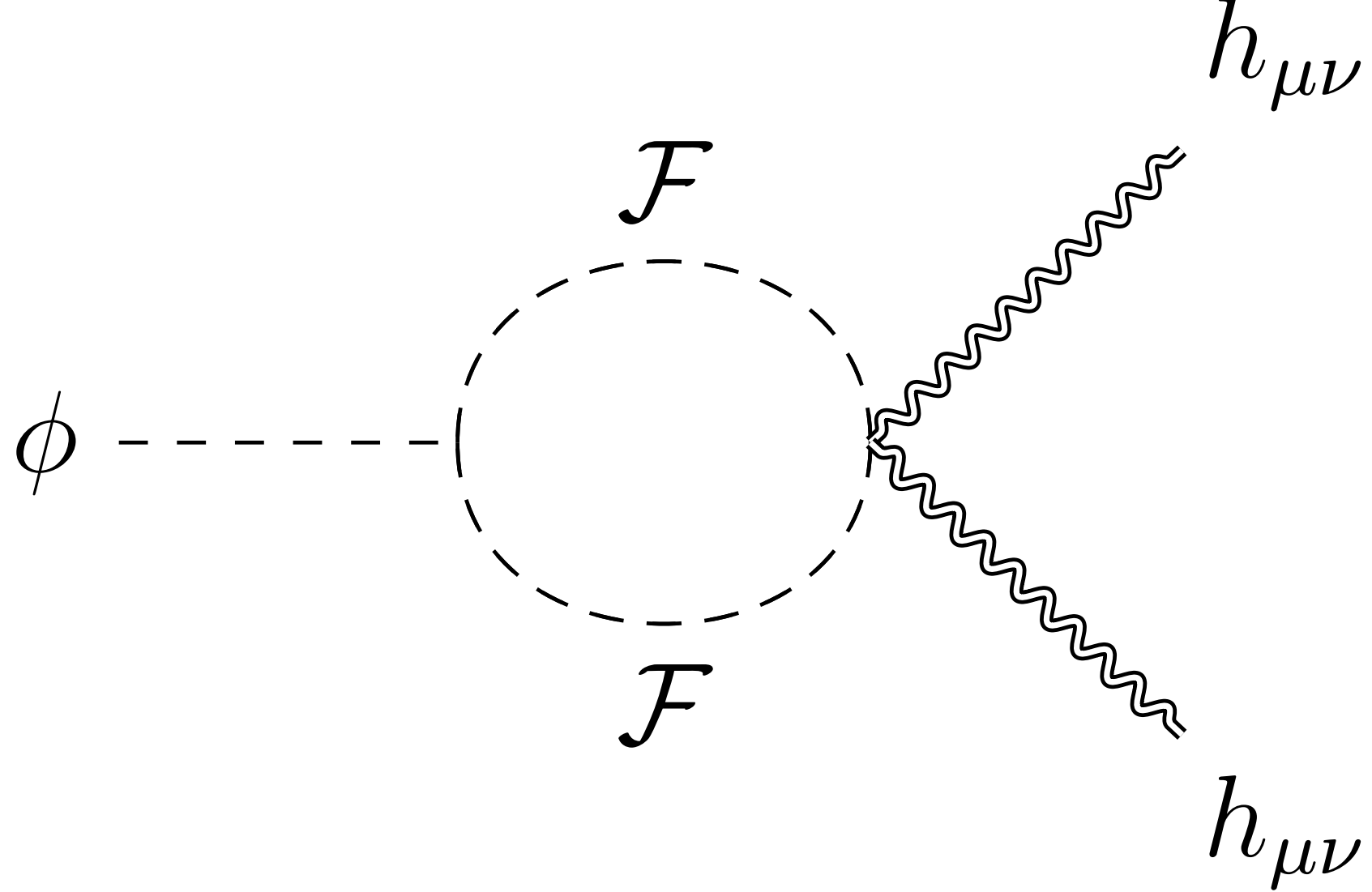}
		\includegraphics[scale=\sepf]{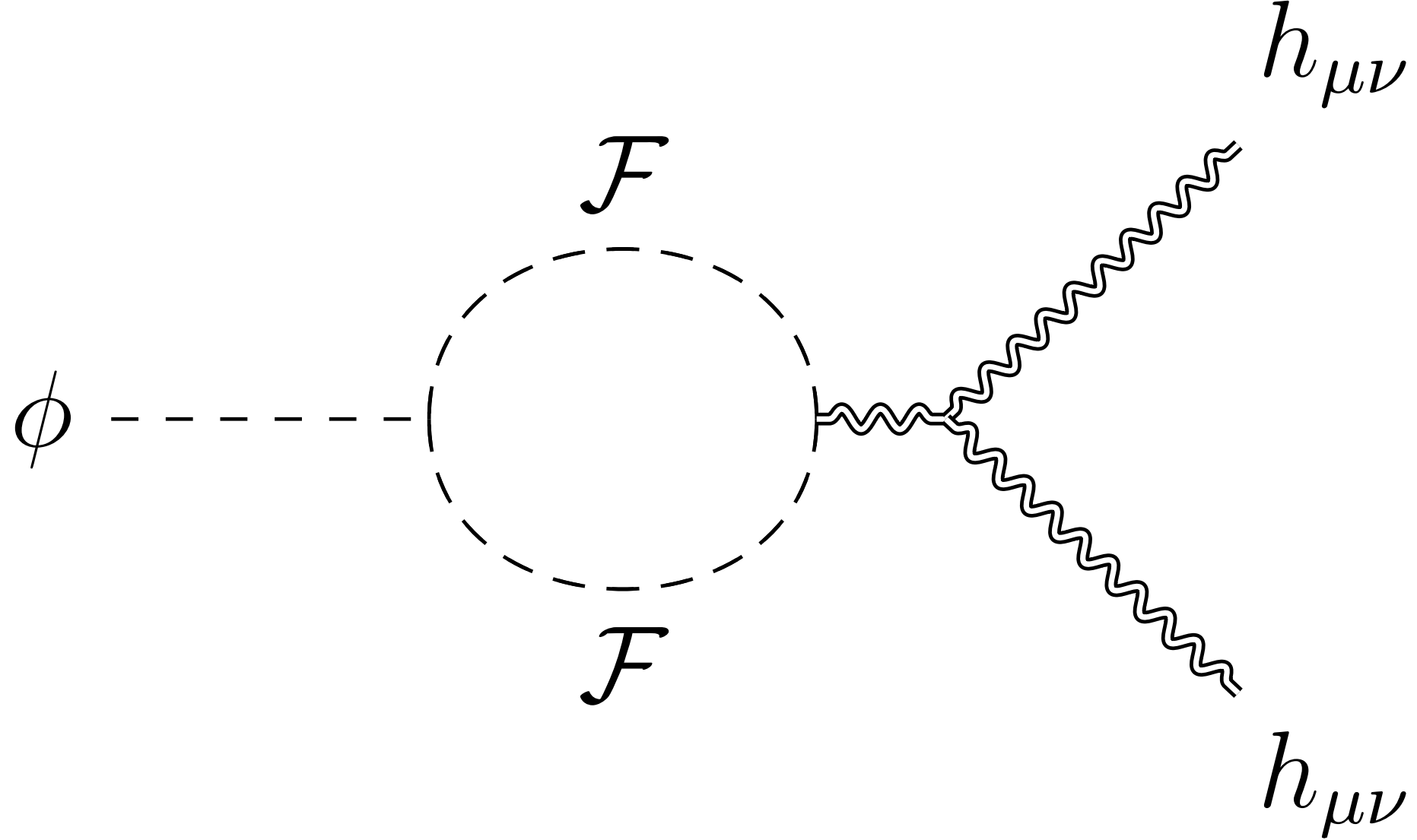}
		\caption {Feynman diagrams for pair production of gravitons from inflaton decay at one-loop level with $\mathcal{F} = \{\psi, \, \varphi \}$. }
		\label{fig:gra_pair_dec}
	\end{figure}

	Along the lines of double graviton production in the previous section, for completeness we note that within our setup with minimal Einstein-Hilbert action, double gravitons can only be sourced via loop-induced inflaton decay. At the one-loop level, the diagrams are shown in Fig.~\ref{fig:gra_pair_dec}. The graviton production rate for a bosonic loop (i.e., $\mathcal{F} = \varphi$) is\footnote{This is similar to Higgs decays into a pair of gravitons \cite{Delbourgo:2000nq}.}
	\begin{align}\label{eq:1-2_rate}
		\Gamma_g^{1 \to 2} \simeq  \frac{3\,m_\phi^3\,  \mu^2}{2048\,\pi^5M_P^4}\,,
	\end{align}
	where $\mu$ denotes the dimensional trilinear coupling\footnote{We note that in polynomial inflation \cite{Drees:2021wgd, Drees:2022aea}, it is possible to have the inflaton running in the loop due to the presence of a $\phi^3$ term.}. We note that the loop-induced inflaton decay as shown in Eq.~\eqref{eq:1-2_rate} is smaller compared to the inflaton annihilation in Eq.~\eqref{eq:2-2_inflaton} as $\rho_\phi > m_\phi^2 \mu^2$. We will show this more explicitly when we compare the GW spectra. For a fermionic loop, the rate scales as $\Gamma_g^{1 \to 2} \propto m_\phi^3 m_\psi^2/(M_P^4)$, which vanishes for a massless fermion in the loop. 
	\section{Graviton Energy  Spectrum}\label{sec:energy_spectrum}
	With the graviton production rates at hand, we are now ready to discuss the graviton energy spectrum, which will be used to obtain the GW spectrum. We need to first discuss the differential energy spectrum, which is defined via
	\begin{align}\label{eq:dGWdE}
		\frac{d\rGW}{dE_\omega} & \equiv E_\omega \frac{dn_g}{dE_\omega} \,,
	\end{align}
	where $n_g$ denotes the graviton number density, $\rGW$ the total GW energy density, and $E_\omega$ the graviton energy. The total energy density stored in GWs can be obtained by integrating Eq.~\eqref{eq:dGWdE} over $E_\omega$, namely 
	\begin{align}
		\rGW = \int  \frac{d\rGW}{dE_\omega}  dE_\omega\,.
	\end{align}
	The evolution for $\rGW $ follows the Boltzmann equation:
	\begin{align}\label{eq:Bol_rhog}
		\frac{d \rGW}{dt} + 4 H\,\rGW= \rp\,\Gamma_g \,,
	\end{align}
	where $\Gamma_g$ corresponds to the graviton production rate discussed in the previous section. The evolution of the graviton number density $n_g$ is governed by the Boltzmann equation:
	\begin{align}\label{eq:Bol_ng}
		\frac{dn_g}{dt} + 3 H n_g= \frac{\rp}{m_\phi} \Gamma_g\,.
	\end{align}
	As a way of cross-checking for consistency, we note that Eq.~\eqref{eq:Bol_ng} can reproduce Eq.~\eqref{eq:Bol_rhog} by using $\rGW  = n_g E_\omega$ and $dE_\omega/dt = -H E_\omega$.
	
	\subsection{\texorpdfstring{$1\to 3$}{1 to 3} 
		Bremsstrahlung}
	
	For gravitons produced from  $1\to 3$ Bremsstrahlung, we note that there is already a spectrum {\it at production} with $0 < \Eom < m_\phi /2$. Thereafter, in order to compute the differential spectrum, one shall rewrite  Eq.~\eqref{eq:Bol_ng} in a differential form, which is
	\begin{align}\label{eq:Bol_ng_diff}
		\frac{d}{dt} \frac{dn_g}{d\Eom}+ 3 H \frac{dn_g}{d\Eom}= \frac{\rp}{m_\phi} \frac{d\Gamma_g^{1 \to 3}}{d\Eom}\,.
	\end{align}
	We note that by integrating over $\Eom$ for Eq.~\eqref{eq:Bol_ng_diff}, one reproduces Eq.~\eqref{eq:Bol_ng}. Using the definition of the differential energy density in Eq.~\eqref{eq:dGWdE}, it follows that \cite{Barman:2023rpg}
	\begin{align}\label{eq:Bol_rGW_diff}
		\frac{d}{dt} \frac{d\rGW}{d\Eom}+ 4 H \frac{d\rGW}{d\Eom}= \rp \frac{\Eom}{m_\phi} \frac{d\Gamma_g^{1 \to 3}} {d\Eom} \,.
	\end{align}
	Note that in Eq.~\eqref{eq:Bol_rGW_diff}, the term $\Eom/m_\phi$ denotes the fraction of inflaton energy goes to GW in each of the $1 \to 3 $ decay. The rest part of the inflaton energy goes to radiation $\rR$, which dilutes the generated GWs. We remind the reader again that the gravitons carry an energy $0 < \Eom < m_\phi /2$ {\it at production}, and then such energy bin gets redshifted till the end of reheating with energy $\Eom(\Trh) \equiv (a_p/\arh) \Eom $ with $a_p$ denoting the scale factor when the gravitons are produced.\footnote{From Eq.~(3.9), it follows that $\rho_R \propto a^{-3/2}$, which leads to the scaling of the temperature $T \propto a^{-3/8}$ during reheating. Consequently, we have $\left(\frac{a_p}{a_{\text{rh}}}\right) \propto \left(\frac{T_p}{T_{\text{rh}}}\right)^{-8/3}$, where $T_p$ denotes the temperature at $a = a_p$.}  Consequently, low energy gravitons with energy $0< \Eom(\Trh) \leq m_\phi /2 \left(\Tmax/\Trh\right)^{-8/3}$ can be produced throught the reheating phase, and higher energy graviton with energy $m_\phi /2 \left(\Tmax/\Trh\right)^{-8/3}  \leq \Eom(\Trh) \leq m_\phi /2 $ can be produced in later phase of reheating. 
	
	Taking into account the dilution as well as the redshift effects, the solutions for the spectrum at the end of reheating can be obtained by solving Eq.~\eqref{eq:Bol_rGW_diff}. We refer to Ref.~\cite{Barman:2023rpg} for more  details. The full spectrum is presented in Appendix \ref{sec:full_spectrum}. In the regime with low energy $\Eom(\Trh)  \ll m_\phi/2$, it takes a simple form:
	\begin{align}\label{eq:1to3_spectrum}
		\frac{d\rGW(\Trh) }{d \Eom(\Trh)} \simeq \frac{\Trh^2 }{24\, \pi^2\, M_P} \sqrt{\frac{\gs}{10}}  \log \left(\frac{\Tmax}{\Trh}\right) 
		& \begin{cases}
			\mu^2
			& \text{bosonic}\,,\\
			y^2\, m_\phi^2\, & \text{fermionic}\,.
		\end{cases}
	\end{align}
	%
	
	\subsection{\texorpdfstring{$2\to 2$}{2 to 2}  Scattering}\label{sec:spectrum}
	For gravitons produced from $2 \to 2$ scatterings, the graviton energy is fixed to be $m_\phi$ at production, i.e., there is no spectrum at production. This implies we do not need to work on the differential form of the Boltzmann equation. However, depending on when the gravitons are produced, they receive different redshifts, leading to a spectrum at the end of reheating.

	To obtain the spectrum, it is more convenient to rewrite the Boltzmann equation Eq.~\eqref{eq:Bol_ng} using the scale factor $a$ and the comoving graviton number density $N_g \equiv n_g a^3$, leading to
	\begin{align}\label{eq:Bol_Ng}
		\frac{dN_g}{da} = \frac{a^2}{H}      \frac{\rp}{m_\phi} \Gamma_g\,.
	\end{align}
	Assume  a graviton produced at $a = a_p$ during reheating,  the spectrum at the end of reheating (i.e. when $T= \Trh$ or $a=\arh$) is then%
	\begin{align}\label{eq:spectrum_end}
		\frac{dn_g(\Trh)}{d\Eom(\Trh)} 
		&\equiv  \frac{1}{\arh^3}\frac{dN_g(\Trh)}{d\Eom(\Trh)} 
		= \frac{1}{\arh^3}\frac{dN_g(a_p)}{da_p} \frac{da_p}{d\Eom(\Trh)}\,, 
	\end{align}
	with which one can finally obtain the differential energy density:
	\begin{align}\label{eq:spectrum_rh}
		\frac{d\rGW(\Trh)}{dE_\omega(\Trh)}  &=\Eom(\Trh) \frac{dn_g(\Trh)}{d\Eom(\Trh)} \,.
	\end{align}
	Due to redshifts, the graviton at the end of reheating is given by $E_\omega (\Trh) = E_\omega(a_p)  \frac{a_p}{\arh}$, where $a_p$ corresponds to the scale factor at production.  Taking the derivative of $E_\omega (T_{\text{rh}})$ with respect to  $a_p$, we obtain $\frac{d E_\omega (T_{\text{rh}})}{da_p} = \frac{E_\omega (a_p) }{a_{\text{rh}}}$, which implies that $\frac{da_p}{d E_\omega (T_{\text{rh}})} = \frac{a_{\text{rh}}}{E_\omega (a_p) }$. We remind the reader again that the graviton energy is $\Eom(a_p) = m_\phi$ at production for $2 \to 2$ scattering.
	
	Before closing this section, we note that for $1 \to 2$ decay, the formalism presented here also applies. However, there is a factor of two difference as $\Eom(a_p) = m_\phi/2$ at production for $1 \to 2$ decay. Moreover, there is an extra factor of two appearing on the right-hand side of Eq.~\eqref{eq:Bol_Ng} since two gravitons are produced in each decay.
	\subsubsection{Inflaton and Decay Product Scattering}
	Using the gravitons production rates presented in Eq.~\eqref{eq:2-2_rate}, we find the differential spectrum  Eq.~\eqref{eq:spectrum_rh} for inflaton and decay product  $2\to 2$ scattering is
	\begin{align}\label{eq:spectrum_rh_sol}
		\frac{d\rGW(\Trh)}{dE_\omega(\Trh)} \simeq&
		\begin{cases}
			\frac{\mu^2 \, \Trh^4}{48\, \pi^2\, M_P\,m_\phi^2} \sqrt{\frac{\gs}{10}}  \left[\frac{E_\omega(\Trh)}{m_\phi} \right]^{3/4} &\text{inflaton-boson},
			\\
			\frac{3\,y^2\, \Trh^6}{4\, \pi^2\, M_P\, m_\phi^2} \sqrt{\frac{\gs}{10}} & \text{inflaton-fermion}.
		\end{cases}
	\end{align}
	The first line corresponds to inflaton scattering with a bosonic decay product, and the second line corresponds to the fermionic case.  As will be shown shortly, the different scaling of graviton energy controls the scaling of the GW spectrum with frequency. 
	By integrating Eq.~\eqref{eq:spectrum_rh_sol} over the graviton energy $m_\phi(\aend/\arh) \leq E_\omega(\Trh) \leq m_\phi$, one can obtain the total energy density $\rGW(\Trh)$ at the end of reheating:
	\begin{align}\label{eq:Rhorh}
		\rGW(\Trh) \simeq&
		\begin{cases}
			\frac{\mu^2\, \Trh^4}{84 \pi^2\, M_P\, m_\phi} \sqrt{\frac{\gs}{10}}    &\text{inflaton-boson},
			\\
			\frac{3\, y^2\, \Trh^6}{4\, \pi^2\, M_P\,  m_\phi} \sqrt{\frac{\gs}{10}}  & \text{inflaton-fermion}.
		\end{cases}
	\end{align}
	As a way of cross checking for consistency, we note that the same results as shown in Eq.~\eqref{eq:Rhorh} have been obtained by directly solving Eq.~\eqref{eq:Bol_rhog}. 
	
	\subsubsection{Inflaton and Inflaton Scattering}
	Using the graviton production rates presented in Eq.~\eqref{eq:2-2_rate}, we find the differential spectrum Eq.~\eqref{eq:spectrum_rh}  for inflaton-inflaton scattering is 
	\begin{align}\label{eq:spectrum_rh_sol_inflaton}
		\frac{d\rGW(\Trh)}{dE_\omega(\Trh)} \simeq&\frac{\gs\, \pi^2\, \Trh^6}{960\,  M_P^3}  \sqrt{\frac{\gs}{10}}  \left[\frac{m_\phi}{E_\omega(\Trh)}\right]^{3/2}  \quad \text{inflaton-inflaton}\,,
	\end{align}
	from which one can also compute the total energy stored in GW by integrating $m_\phi (\aend/\arh)\leq \Eom(\Trh) \leq m_\phi$, which is
	\begin{align} \label{eq:rho_GW_inflaton}
		\rGW(\arh) = \frac{\sqrt{3}\, m_\phi \, \rp(\aend)^{1/6}\, \Trh^{16/3}}{16\, \pi\, M_P^3} \left(\frac{\gs\, \pi^2}{30}\right)^{4/3}  \quad \text{inflaton-inflaton}\,.
	\end{align}
	Again, for crosschecking, we note that the result presented in Eq.~\eqref{eq:rho_GW_inflaton} is consistent with Eq.~(22) of Ref.~\cite{Choi:2024ilx}, where the authors have used method to directly solve Eq.~\eqref{eq:Bol_rhog}.

	\subsection{One-loop Induced \texorpdfstring{$1\to 2$}{1 to 2}  Decay}
	With the graviton production rate shown in Eq.~\eqref{eq:1-2_rate}, we obtain  the spectrum  Eq.~\eqref{eq:2-2_rate} for $1 \to 2$ decays is given by
	\begin{align}\label{eq:spectrum_rh_1to2_sol}
		\frac{d\rGW(\Trh)}{dE_\omega(\Trh)} \simeq&\frac{3\,\sqrt{2}\,\mu^2\, m_\phi^2\, \Trh^2}{512\, \pi^4\,  M_P^3}  \sqrt{\frac{\gs}{10}}  \left[\frac{E_\omega(\Trh)}{m_\phi}\right]^{3/2}  \,,
	\end{align}
	from which one can also compute the total energy stored in GW by integrating $m_\phi/2  (\aend/\arh)\leq \Eom(\Trh) \leq m_\phi /2$, which is
	\begin{align} \label{eq:rho_GW_inflaton_decay}
		\rGW(\arh) = \frac{3\,\mu^2\, m_\phi^3\, \Trh^2}{5120\, \pi^4\,  M_P^3}  \sqrt{\frac{\gs}{10}} \,.
	\end{align}
	Once again, we have confirmed that the same result shown in Eq.~\eqref{eq:rho_GW_inflaton_decay} has been obtained by directly solving Eq.~\eqref{eq:Bol_rhog}. The reason we mention two different methods for computing $\rho_{\text{GW}}(\arh)$ three times is to demonstrate that the formalism to compute the differential spectrum presented in Sec.~\ref{sec:spectrum}, which, to the best of our knowledge, has not been shown in the literature in the context of GWs, is robust.
	
	\section{Gravitational Wave Spectrum}\label{sec:GW}
	With the graviton energy spectrum at the end of reheating, we are now ready to compute the present-day GW spectrum. The primordial GW spectrum at present, $\Omega_{\text{GW}}(f)$, per logarithmic frequency $f$, is defined as \cite{Barman:2023rpg}
	\begin{align} \label{eq:oGW}
		\oGW(f) = \frac{1}{\rho_c}\, \frac{d\rGW}{d\ln f} = \Omega_\gamma^{0}\, \frac{d(\rGW/\rR)}{d\ln f} = \Omega_\gamma^{0}\, \frac{\gs(\Trh)}{\gs(T_0)} \left[\frac{\gss(T_0)}{\gss(\Trh)}\right]^{4/3}\,\frac{d\left[ \rGW(\Trh)/\rR(\Trh) \right]}{d\ln \Eom(\Trh)}\,,
	\end{align}
	where $\rho_c$ corresponds to the critical energy density and $\Omega_\gamma^{0} h^2 \simeq 2.47\cdot10^{-5}$  photon abundance at present~\cite{Planck:2018vyg}. GW frequency at present $f$ is associated with the graviton energy at the end of reheating  $\Eom (\Trh)$  via 
	\begin{align}\label{eq:f}
		f &\equiv \frac{\Eom (T_0)}{2 \pi} =  \frac{\Eom(\Trh)}{2 \pi}\, \frac{\arh}{a_0} = \frac{\Eom(\Trh)}{2 \pi}\, \frac{T_0}{\Trh} \left[\frac{\gss(T_0)}{\gss(\Trh)}\right]^{1/3}\,,
	\end{align}
	where, in the second step, we account for the redshift of the graviton energy from the end of reheating until today, using the scale factor \( a_0 \). Finally, in the last step, we apply entropy conservation, which allows us to express the ratio of scale factors as the ratio of temperatures and degrees of freedom $\gss$.
	\subsection{\texorpdfstring{$1\to 3$}{1 to 3} Bremsstrahlung}
	For gravitons produced from $1\to 3$ Bremsstrahlung, the corresponding  GWs have been  recently investigated in Refs.~\cite{Barman:2023ymn, Barman:2023rpg}. 
	Using Eq.~\eqref{eq:1to3_spectrum}, we find that the spectrum takes a form:
	\begin{align}\label{eq:oGW2_1to3}
		\oGW^{1\to3}h^2(f)   &   \simeq  8.3  \cdot 10^{-20}  \cdot \log \left(\frac{\Tmax}{\Trh}\right) \times  \nonumber \\
		&
		\begin{cases}
			\left( \frac{\mu}{10^{11}~\text{GeV}}\right)^{2} \left( \frac{10^{13}~\text{GeV}}{\Trh}\right)   \left( \frac{f}{10^{9}~\text{Hz}}\right)  &\text{bosonic decay},
			\\
			\left( \frac{y}{10^{-2}}\right)^{2} \left( \frac{m_\phi}{10^{13}~\text{GeV}}\right)^2  \left( \frac{10^{13}~\text{GeV}}{\Trh}\right)   \left( \frac{f}{10^{9}~\text{Hz}}\right) & \text{fermionic decay}\,,
		\end{cases}
	\end{align}
	which fits well the full spectrum  before the peak as shown in Fig.~\ref{fig:OGW_1to3_compare} in Appendix~\ref{sec:full_spectrum}. The maximum temperature $\Tmax$ is given in Eq.~\eqref{eq:Tmax2}, and 
	$\log \left(\frac{\Tmax}{\Trh}\right) \sim \mathcal{O}(1)$ for $\Tmax \gg \Trh$.
	The frequency $f$ for GW from  $1\to 3$ Bremsstrahlung, satisfies
	\begin{equation}\label{eq:fpeak}
		f \lesssim f_{\text{peak}}  \simeq  \frac{m_\phi}{4\pi}\, \frac{T_0}{\Trh} \left[\frac{\gss(T_0)}{\gss(\Trh)}\right]^{1/3} \simeq 9.5 \cdot10^{9}\,\left( \frac{m_\phi}{10^{13}~\text{GeV}}\right)\, \left(\frac{ 10^{13}~\text{GeV}}{\Trh}\right) \text{Hz}\,,
	\end{equation}
	where we have used $\gss(T_0) = 3.94$ and $\gss(\Trh) = 106.75$. Such bound comes from the fact that the energy of emitted graviton $E_\omega$ could be at most half of the inflaton mass, namely $E_\omega \leq m_\phi/2$ during reheating. 
	Note that the spectrum peaks at a frequency $f_{\text{peak}} \simeq 9.5 \times 10^9~\text{Hz}\left(\frac{m_\phi}{\Trh}\right) \gtrsim 9 \times 10^9~\text{Hz}$ if $m_\phi \gtrsim \Trh$. The dependence of the Bremsstrahlung GWs on the inflaton mass $m_\phi$ as well as the reheating temperature $\Trh$  makes it possible to probe the reheating parameters with future experiments. We demonstrate this proposal in Appendix \ref{sec:probing}.

	Before closing this section, we note that gravitons could be produced after reheating with less redshift (compared to those generated during reheating) until today, thereby leading to higher frequencies $f > f_{\text{peak}}$. However, after reheating, the inflaton energy density scales as $\rp(a) \propto e^{-(a/\arh)^{2}}$, as seen in the first line of Eq.~\eqref{eq:rhophi_sol}.\footnote{We remind the reader that after reheating, the Hubble parameter scales as $H = H(\arh) (a/\arh)^2$ in a radiation phase.} Consequently, although higher-frequency gravitons can be generated after reheating, the corresponding GW amplitude is exponentially suppressed by a factor $\sim e^{-(f/f_{\text{peak}})^{2}}$.
	\subsection{\texorpdfstring{$2\to 2$}{2 to 2}  Scattering}
	In this section, we present the GW spectrum for $2\to 2$ scatterings.
	\subsubsection{Inflaton and Decay Product Scattering}
	By using Eq.~\eqref{eq:spectrum_rh_sol}, we find the GW spectrum for $2\to 2$ scattering between inflaton and its decay product is given by
	\begin{align}\label{eq:2-2_spectrum}
		\oGW^{2\to2}h^2(f)   &   \simeq \nonumber \\
		&
		\begin{cases}
			4.6\cdot 10^{-21} \left( \frac{\mu}{10^{11}~\text{GeV}}\right)^{2} \left( \frac{\Trh}{10^{13}~\text{GeV}}\right)^{7/4}  \left( \frac{10^{13}~\text{GeV}}{m_\phi}\right)^{11/4}  \left( \frac{f}{10^{9}~\text{Hz}}\right)^{7/4}  &\text{inflaton-boson},
			\\
			1.5\cdot 10^{-18} \left( \frac{y}{10^{-2}}\right)^{2} \left( \frac{\Trh}{10^{13}~\text{GeV}}\right)^{3}  \left( \frac{10^{13}~\text{GeV}}{m_\phi}\right)^{2}  \left( \frac{f}{10^{9}~\text{Hz}}\right) & \text{inflaton-fermion},
		\end{cases}
	\end{align}
	where the different scaling on $f$ arises from the distinct scaling of the graviton energy $E_\omega(\Trh)$ in the spectrum Eq.~\eqref{eq:spectrum_rh_sol}, which is controlled by 
	production rates as presented in  Eq.~\eqref{eq:2-2_rate}. Note that the upper bound of the frequency $f$ from $2\to 2$ scattering is twice larger compared to that presented in Eq.~\eqref{eq:fpeak}. There is also a lower bound on the frequency as explained as follows. Note that the earliest produced graviton at $a = \aend$ receives most redshifts, leading to a graviton energy at the end of reheating to be $\Eom(\Trh) = m_\phi (\aend /\arh) = m_\phi \left(\Tmax/\Trh\right)^{-8/3}$. To sum up, the frequency for GW from $2 \to 2 $ scattering is
	\begin{equation} \label{eq:f_range_2to2}
		f_{1}\leq f \lesssim  f_{2}\,,
	\end{equation}
	where
	\begin{align}
		& f_{1} \simeq  1.9 \cdot10^{2}\,\left( \frac{m_\phi}{10^{13}~\text{GeV}}\right)\, \left(\frac{ 10^{13}~\text{GeV}}{\Trh}\right) \left( \frac{\Tmax/\Trh} {10^{3}}\right)^{-8/3}\text{Hz}\,, \label{eq:f1} \\
		&f_{2}\simeq 1.9 \cdot10^{10}\,\left( \frac{m_\phi}{10^{13}~\text{GeV}}\right)\, \left(\frac{ 10^{13}~\text{GeV}}{\Trh}\right) \text{Hz} \label{eq:f2}\,.
	\end{align}
	Note that the larger $\Tmax/\Trh$ is, the smaller $f_{1}$ can be.   It is important to note that GW with $f > f_{2}$ could be produced after reheating, but the corresponding GW spectrum is exponentially suppressed by a factor $\sim e^{-(f/f_{2})^{2}}$, similar to the discussion in previous subsection.
	\subsubsection{Inflaton Inflaton Scattering}
	For inflaton inflaton scattering, after using Eq.~\eqref{eq:spectrum_rh_sol_inflaton} we find 
	\begin{align}\label{eq:2-2_spectrum_inflaton}
		\oGW^{2\to2}h^2(f)   &   \simeq 
		3 \cdot 10^{-22} \left( \frac{\Trh}{10^{13}~\text{GeV}}\right)^{3/2}  \left( \frac{m_\phi}{10^{13}~\text{GeV}}\right)^{3/2}  \left( \frac{10^{9}~\text{Hz}}{f}\right)^{1/2}  &\text{inflaton-inflaton},
	\end{align}
	where the frequency is also determined by Eq.~\eqref{eq:f_range_2to2}.  We note that the scaling of the spectrum on $f$, $m_\phi$ as well as $\Trh$ is consistent with that presented in Ref.~\cite{Choi:2024ilx}. Similar to the previous case, the spectrum  with $f > f_{2}$  is exponentially suppressed by a factor $\sim e^{-(f/f_{2})^{2}}$.
	\subsection{One-loop Induced \texorpdfstring{$1\to 2$}{1 to 2}  Decay}
	Finally, for the loop induced GW, with Eq.~\eqref{eq:spectrum_rh_1to2_sol} we find  
	\begin{align}\label{eq:1-2_spectrum}
		\oGW^{1\to 2}h^2(f)   &   \simeq
		3.5 \cdot 10^{-34} \left( \frac{\mu}{10^{11}~\text{GeV}}\right)^{2} \left( \frac{\Trh}{10^{13}~\text{GeV}}\right)^{1/2}  \left( \frac{m_\phi}{10^{13}~\text{GeV}}\right)^{1/2}  \left( \frac{f}{10^{9}~\text{Hz}}\right)^{5/2}\,.
	\end{align}

	The lower and upper limits of the frequency are half of those from the previous $2 \to 2$ case in Eq.~\eqref{eq:f_range_2to2}, since the graviton energy at production from decay is $m_\phi/2$. Again, the spectrum with higher frequencies is exponentially suppressed by a factor $\sim e^{-(2\,f/f_{2})^{2}}$ for $f > f_2/2$.
	
	\subsection{Results}
	\begin{figure}[t!]
		\def\sepf{0.6}
		\centering
		\includegraphics[scale=\sepf]{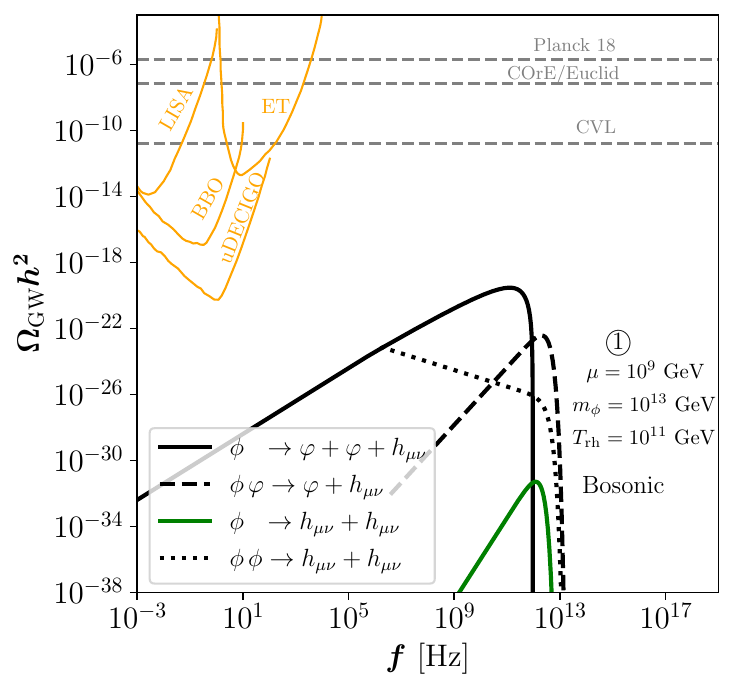}
		\includegraphics[scale=\sepf]{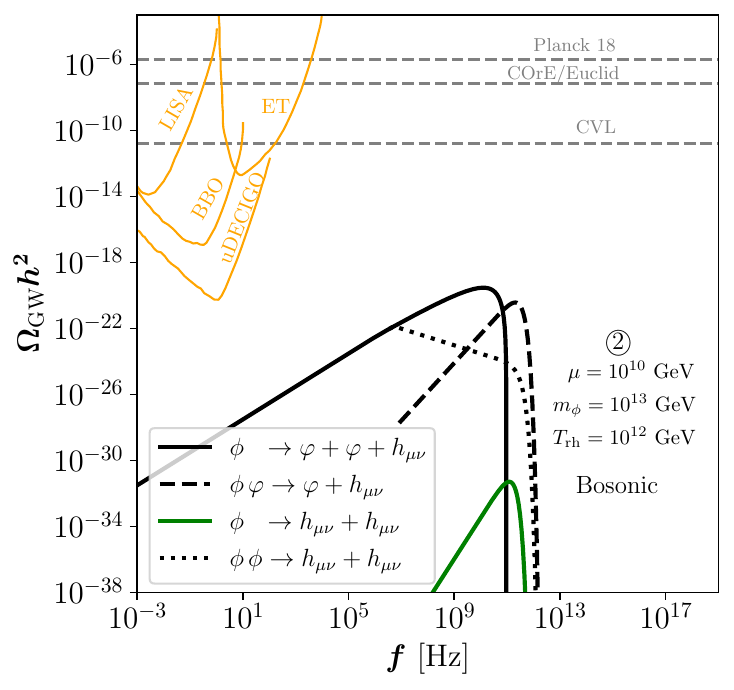}
		\includegraphics[scale=\sepf]{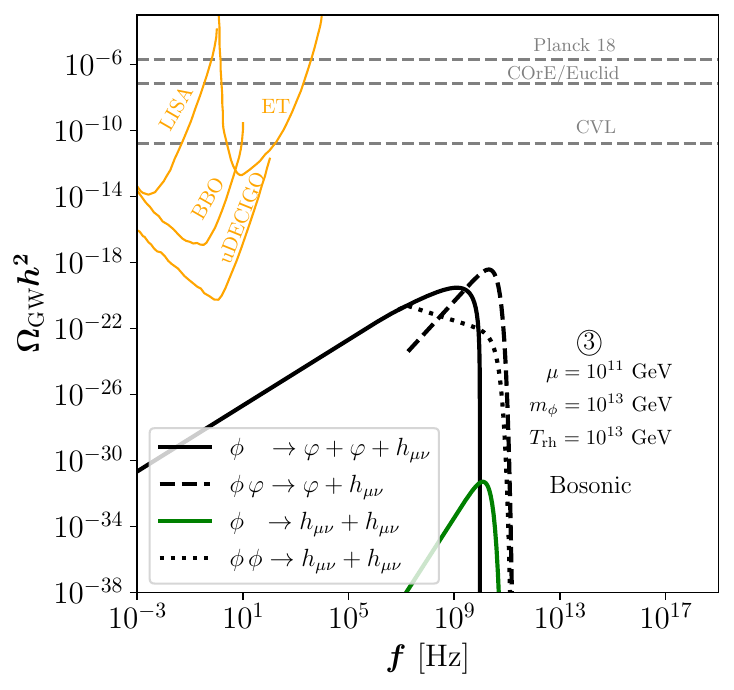}
		\includegraphics[scale=\sepf]{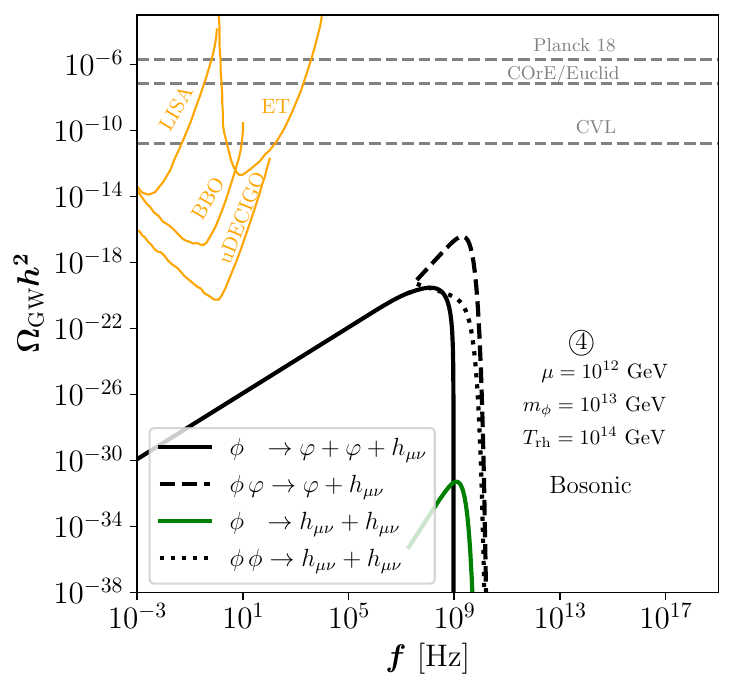}
		\caption{GW from Bremsstrahlung (black solid line), inflaton and decay product scattering (black dashed line), one-loop induced inflaton decay (green line), and inflaton-inflaton scattering (black dotted line) for bosonic processes with four sets of benchmark model parameters.
		}
		\label{fig:OGW_compare_Bos_n=2}
	\end{figure} 

	In this section, we present the GW spectrum based on the previous subsections, focusing on comparing spectra from different sources. The objective is to identify the dominant process and the corresponding conditions under which it occurs.

	\subsubsection{Bosonic Reheating}
	In Fig.~\ref{fig:OGW_compare_Bos_n=2}, we compare the spectra for various processes: $1 \to 3$ Bremsstrahlung (black solid line), $2 \to 2$ scattering between inflaton and decay products (black dashed line), one-loop induced inflaton decay (green line), and inflaton-inflaton scattering (black dotted line) in bosonic processes. Additionally, we include several proposed high-frequency GW detectors such as LISA~\cite{LISA:2017pwj}, the Einstein Telescope (ET)~\cite{Punturo:2010zz, Hild:2010id, Sathyaprakash:2012jk, Maggiore:2019uih}, the Big Bang Observer (BBO)~\cite{Crowder:2005nr, Corbin:2005ny, Harry:2006fi}, and ultimate DECIGO (uDECIGO)~\cite{Seto:2001qf, Kudoh:2005as}. The energy stored in GWs exhibits characteristics similar to dark radiation, contributing to the effective number of neutrino species, denoted as $ N_{\text{eff}} $ ~\cite{Caprini:2018mtu}. The Planck 2018 mission provides a 95\% confidence level (CL) result of $ N_{\text{eff}} = 2.99 \pm 0.34 $~\cite{Planck:2018vyg}. Future experiments like COrE~\cite{COrE:2011bfs} and Euclid~\cite{EUCLID:2011zbd} are expected to significantly improve these constraints at the $ 2\sigma $ level, resulting in $\Delta N_{\text{eff}} \lesssim 0.013 $. Furthermore, Ref.~\cite{Ben-Dayan:2019gll} reports a bound of $ \Delta N_{\text{eff}} \lesssim 3 \times 10^{-6} $ based on a hypothetical cosmic-variance-limited (CVL) CMB polarization experiment. We have considered four sets of benchmark model parameters:
	\begin{itemize}
		\item $\circled{1}$ $\mu = 10^{9}~\text{GeV}$, $m_\phi = 10^{13}~\text{GeV}$, $\Trh = 10^{11}~\text{GeV}$ (upper left),
		\item $\circled{2}$ $\mu = 10^{10}~\text{GeV}$, $m_\phi = 10^{13}~\text{GeV}$, $\Trh = 10^{12}~\text{GeV}$ (upper right),
		\item $\circled{3}$ $\mu = 10^{11}~\text{GeV}$, $m_\phi = 10^{13}~\text{GeV}$, $\Trh = 10^{13}~\text{GeV}$ (lower left),
		\item $\circled{4}$ $\mu = 10^{12}~\text{GeV}$, $m_\phi = 10^{13}~\text{GeV}$, $\Trh = 10^{14}~\text{GeV}$ (lower right).
	\end{itemize}
	These model parameters can be realized, for instance, in large-field polynomial inflation~\cite{Drees:2022aea}. Note that the values of reheating temperature  are related to the inflaton mass and coupling via Eq.~\eqref{trh}. The maximum temperature $\Tmax$ is given by Eq.~\eqref{eq:Tmax2}.
	We have fixed the inflationary tensor-to-scalar ratio $r = 0.01$. A larger (smaller) $r$ corresponds to a larger (smaller) $\Tmax$. We note that changing $\Tmax$ does not significantly affect the amplitude of the Bremsstrahlung GW spectrum as it only depends on $\log(\Tmax / \Trh)$ (cf. Eq.~\eqref{eq:oGW2_1to3}). For GWs from $2 \to 2$ scattering and $1 \to 2$ decay, increasing (decreasing) $\Tmax$ decreases (increases) the value of the lower limit of the GW frequency $f_{\text{1}}$.
	
	For $m_\phi \gtrsim \Trh$, as shown in the upper panels as well as lower left panel of Fig.~\ref{fig:OGW_compare_Bos_n=2}, we find that the spectrum is dominated by Bremsstrahlung (black solid) at frequencies $f < f_{\text{peak}}$, where $f_{\text{peak}}$ is given in Eq.~\eqref{eq:fpeak}. For $f > f_{\text{peak}}$, the GWs from Bremsstrahlung are suppressed, and the spectrum is instead dominated by contributions from scatterings and decays, where scattering between the inflaton and its decay products (black dashed) can be the dominant one. GWs from one-loop induced inflaton decays (green solid) are typically suppressed compared to inflaton-inflaton scattering (black dotted).
	
	In the opposite limit with $\Trh > m_\phi$, corresponding to the lower right panel of Fig.~\ref{fig:OGW_compare_Bos_n=2}, we demonstrate that inflaton and decay product $2 \to 2$ scattering (black dashed) dominates the spectrum in the regime $f > f_{1}$ (with $f_1$ given in Eq.~\eqref{eq:f1}). Regarding Bremsstrahlung and inflaton-inflaton scatterings, we note that the latter gradually becomes comparable to Bremsstrahlung in the regime $f_1 < f < f_{\text{peak}}$. For GWs from one-loop induced inflaton decays, it remains suppressed compared to inflaton-inflaton scattering.
	
	\subsubsection{Fermionic Reheating}
\begin{figure}[!ht]
		\def\sepf{0.6}
		\centering
		\includegraphics[scale=\sepf]
		{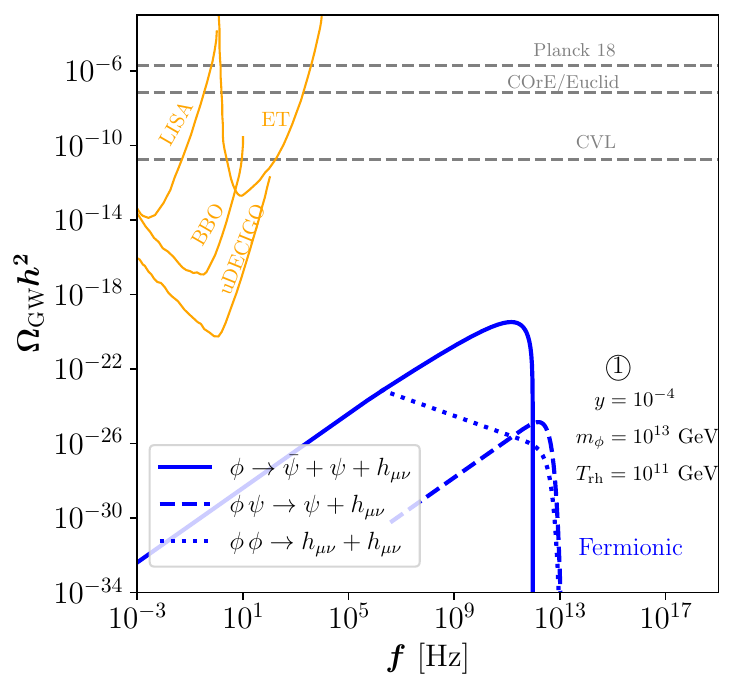}
		\includegraphics[scale=\sepf]
		{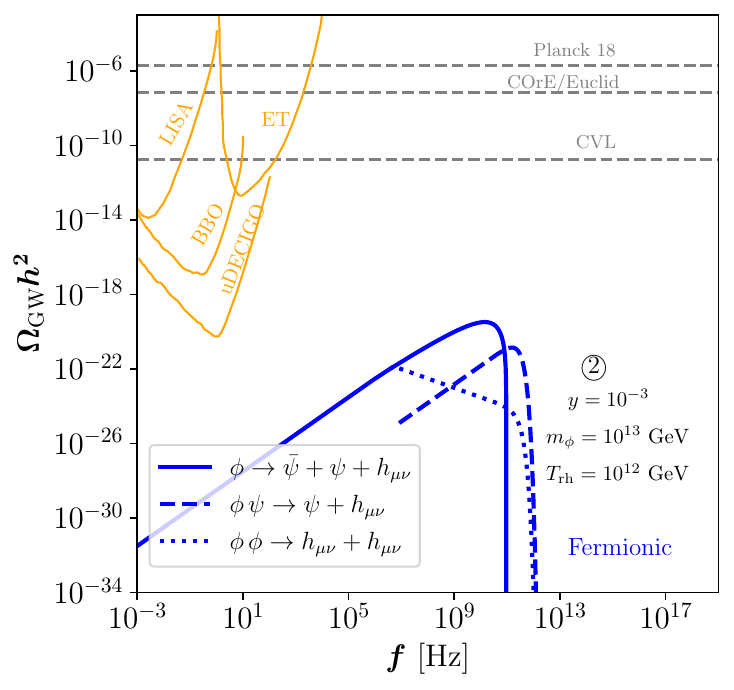}
		\includegraphics[scale=\sepf]
		{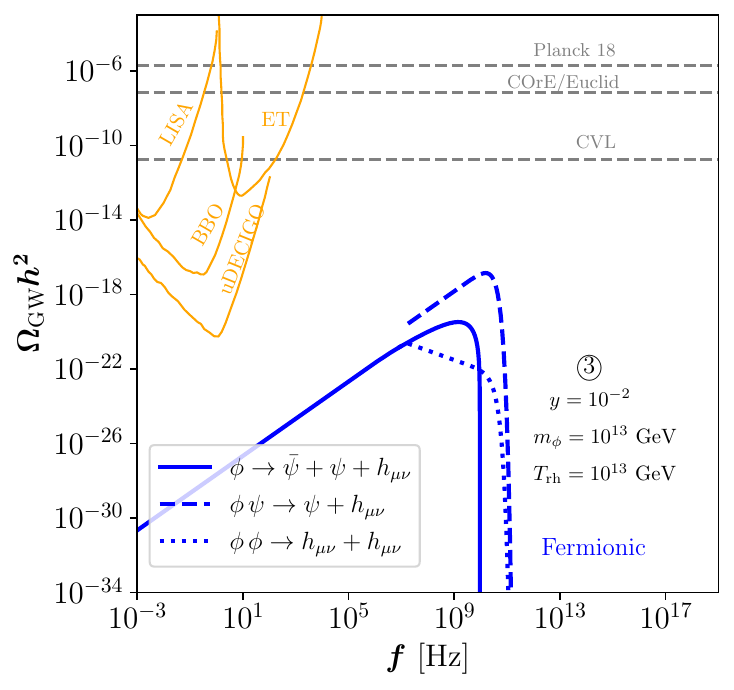}
		\includegraphics[scale=\sepf]
		{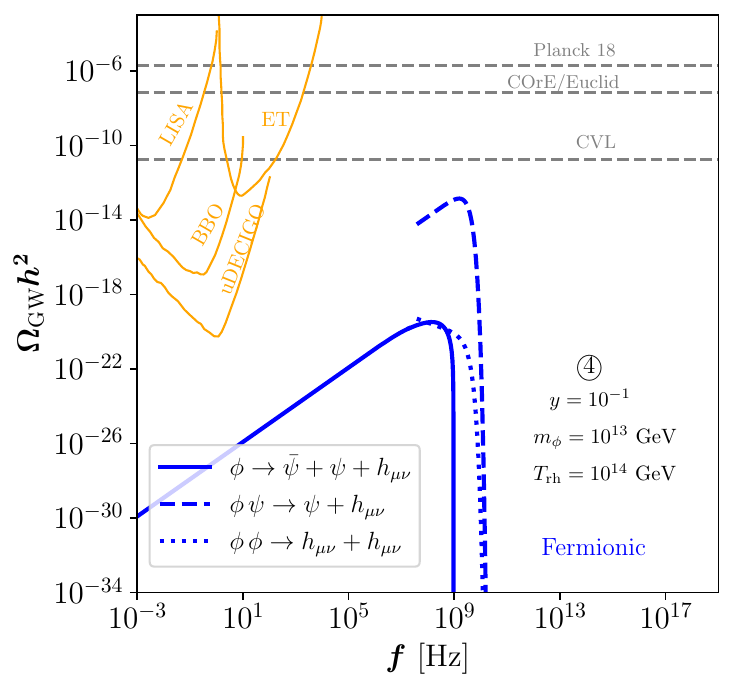}
		\caption{GW from  Bremsstrahlung (blue solid line), inflaton and decay product  scattering (blue  dashed line), and inflaton inflaton scattering (blue dotted line) for fermionic processes with four sets of  benchmark model parameters.}
		\label{fig:OGW_compare_Ferm_n=2}
	\end{figure}

	In Fig.~\ref{fig:OGW_compare_Ferm_n=2}, we present a comparison of the spectra for $1 \to 3$ (blue solid), $2 \to 2$ for inflaton and its decay products (blue dashed), and $2 \to 2$ for inflaton and inflaton scattering (blue dotted) in fermionic processes. We consider four sets of benchmark model parameters\footnote{A large Yukawa coupling could potentially spoil the inflationary predictions when loop corrections to the inflaton potential are included; however, this could be avoided in a supersymmetric setup~\cite{Ellis:2015pla}.}: 
	\begin{itemize}
		\item $\circled{1}$ $y = 10^{-4}$, $m_\phi = 10^{13}~\text{GeV}$, $\Trh = 10^{11}~\text{GeV}$ (upper left),
		\item $\circled{2}$ $y = 10^{-3}$, $m_\phi = 10^{13}~\text{GeV}$, $\Trh = 10^{12}~\text{GeV}$ (upper right),
		\item $\circled{3}$ $y = 10^{-2}$, $m_\phi = 10^{13}~\text{GeV}$, $\Trh = 10^{13}~\text{GeV}$ (lower left),
		\item $\circled{4}$ $y = 10^{-1}$, $m_\phi = 10^{13}~\text{GeV}$, $\Trh = 10^{14}~\text{GeV}$ (lower right).
	\end{itemize} 
	Similar to the bosonic case, we find that when $m_\phi > \Trh$, $1 \to 3$ Bremsstrahlung dominates the spectrum for $f < f_{\text{peak}}$, while $2 \to 2$ scattering dominates at the high-frequency tail, as shown in the upper two panels of Fig.~\ref{fig:OGW_compare_Ferm_n=2}. For $\Trh \gtrsim m_\phi$, $2 \to 2$ inflaton scattering with the decay product  can dominate over $1 \to 3$ Bremsstrahlung in the regime $f > f_{1}$, as shown in the lower panels of Fig.~\ref{fig:OGW_compare_Ferm_n=2}. For $\Trh \sim m_\phi$, unlike the previous bosonic case, $2 \to 2$ inflaton and its decay products scattering is dominant compared to $1 \to 3$ Bremsstrahlung for $f < f_{\text{peak}}$. The differences can be observed in the lower left panel of Fig.~\ref{fig:OGW_compare_Bos_n=2} and Fig.~\ref{fig:OGW_compare_Ferm_n=2}. To facilitate a clearer comparison, we will include a combined figure that incorporates both the bosonic and fermionic cases in the next subsection.
	\subsubsection{Comparison}
\begin{figure}[!ht]
		\def\sepf{0.55}
		\centering
		\includegraphics[scale=\sepf]
		{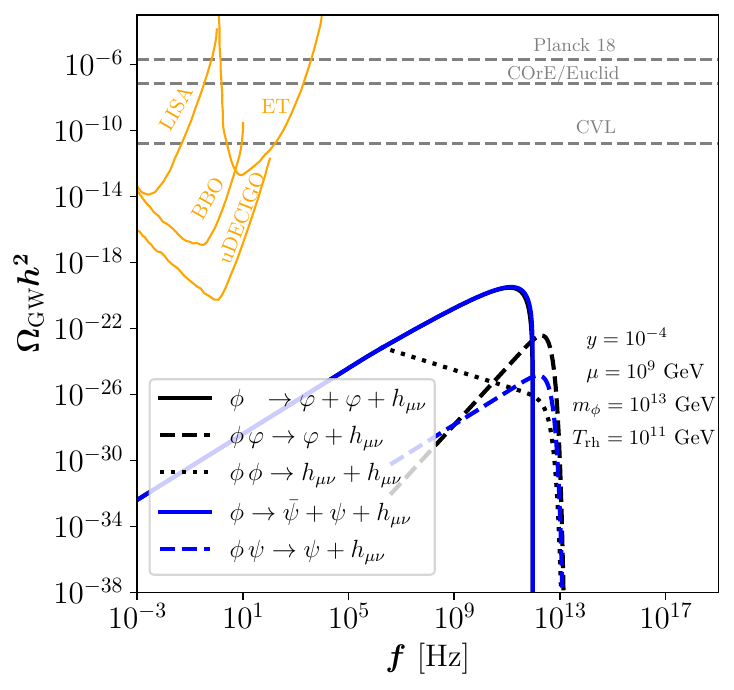}
		\includegraphics[scale=\sepf]
		{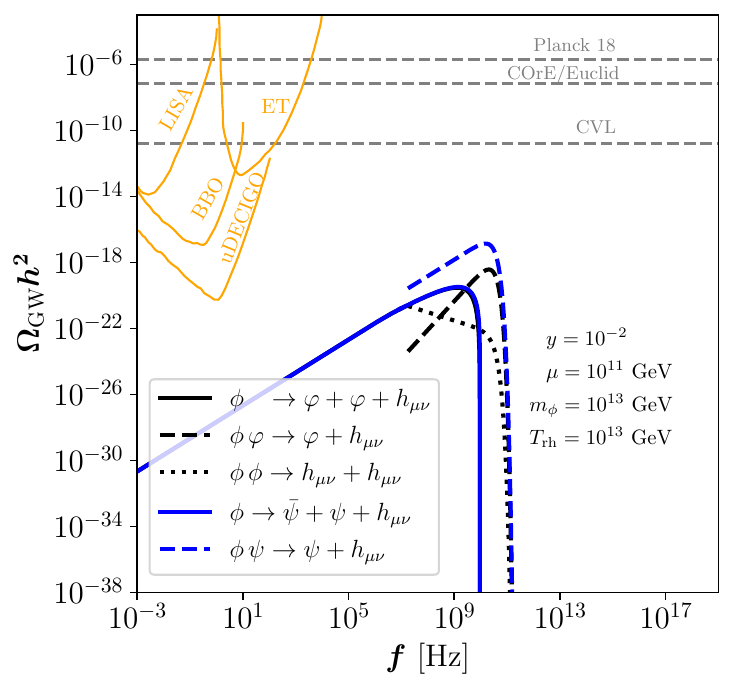}
		\caption{Comparison of GW in for bosonic (black) and fermionic processes (blue) for two sets of benchmark model parameters: $y = 10^{-4}\,, \mu=10^{9}~\text{GeV}\,, m_\phi = 10^{13}~\text{GeV}\,, \Trh = 10^{11}~\text{GeV}$ (left), and  $y = 10^{-2}\,, \mu=10^{11}~\text{GeV}\,, m_\phi = 10^{13}~\text{GeV}\,, \Trh = 10^{13}~\text{GeV}$ (right).}
		\label{fig:OGW_compare_Ferm_Bos}
	\end{figure}

	In Fig.~\ref{fig:OGW_compare_Ferm_Bos}, we compare the GW spectra from bosonic and fermionic processes using two sets of benchmark model parameters. The left panel corresponds to $y = 10^{-4}\,, \mu=10^{9}~\text{GeV}\,, m_\phi = 10^{13}~\text{GeV}\,, \Trh = 10^{11}~\text{GeV}$, and the right panel shows $y = 10^{-2}\,, \mu=10^{11}~\text{GeV}\,, m_\phi = 10^{13}~\text{GeV}\,, \Trh = 10^{13}~\text{GeV}$. 
	
	We note that the Bremsstrahlung GW spectrum is identical in both the bosonic and fermionic cases due to the equivalence in $\Gamma_\phi$ arising from the values of $\mu$ and $y$, with which the GW spectrum depends solely on $\Trh$, $m_\phi$, and $\Tmax$. This explains why the black solid and blue solid lines overlap with each other. 
	Moreover, we find that the GW spectrum for inflaton and its decay product scattering in the bosonic case (black dashed) can be larger than that from the fermionic case (blue dashed)  when $m_\phi > \Trh$. However, the opposite conclusion holds if $m_\phi \lesssim \Trh$. These differences stem from the graviton production rates presented in Eq.~\eqref{eq:2-2_rate}: when $m_\phi > \Trh$, graviton production is more efficient for inflaton scattering with a  bosonic decay product. Conversely, the fermionic case is more efficient when $m_\phi \lesssim \Trh$.

	%
	
	\section{Conclusions}\label{sec:conclusion}
	In this study, we systematically investigate the  ultra-high frequency gravitational waves (GWs) from gravitons generated during inflationary reheating. The processes under consideration include: $(i)$ $1 \to 3$ graviton Bremsstrahlung (Fig.~\ref{fig:diagram}), $(ii)$ $2 \to 2$ scattering of the inflaton and its decay products (Fig.~\ref{fig:2-2diagram}), $(iii)$ pure inflaton $2 \to 2$ scattering (Fig.~\ref{fig:gra_pair_ann}), and $(iv)$ one-loop induced inflaton decay (Fig.~\ref{fig:gra_pair_dec}). We pay particular attention to processes $(ii)$ and $(iv)$, which have not been analyzed in detail. Additionally, we have conducted a comprehensive comparison among the four sources of GWs.
	
	For the $2 \to 2$ inflaton scattering  with  its decay product, despite involving the same couplings as those in the Bremsstrahlung $1 \to 3$ process, we find that the resulting GW spectra exhibit distinct characteristics. We compute the graviton production rate for this process, as presented in Eq.~\eqref{eq:2-2_rate}. We demonstrate that if the reheating temperature exceeds the inflaton mass ($\Trh > m_\phi$), the GW spectrum can be dominated by $2 \to 2$ scattering of the inflaton and its decay product. Conversely, if $\Trh < m_\phi$, the Bremsstrahlung process typically dominates the spectrum before the peak. Additionally, we find  that one-loop induced inflaton $1 \to 2$ decay generates smaller GW signals compared to Bremsstrahlung and pure inflaton $2 \to 2$ scattering. These results are illustrated in Fig.~\ref{fig:OGW_compare_Bos_n=2} and Fig.~\ref{fig:OGW_compare_Ferm_n=2}. In our comparison between the bosonic and fermionic cases, we demonstrate that if $m_\phi \lesssim \Trh$, the GWs from inflaton scattering with fermionic decay product can be larger than those from the bosonic  case. However, when $m_\phi > \Trh$, the GWs from inflaton scattering with bosonic   decay product can be dominant, as shown in Fig.~\ref{fig:OGW_compare_Ferm_Bos}.
	
	In summary, this work provides a comprehensive analysis of non-thermal and unavoidable perturbative sources of ultra-high frequency gravitational waves (GWs) from graviton production during reheating. We have identified the conditions under which dominant sources emerge. 
	
	
	\bibliographystyle{JHEP}
	\section*{Acknowledgments}
	The author wishes to thank B. Barman, M. Becker, N. Bernal, M. Drees, P. Schwaller,  C. Tamarit and O. Zapata  for discussions. YX acknowledges the illuminating discussions and feedback from the Miapbp workshop ``Quantum Aspects of Inflationary Cosmology", where this work was presented. YX has received support from the Cluster of Excellence ``Precision Physics, Fundamental Interactions, and Structure
	of Matter'' (PRISMA$^+$ EXC 2118/1) funded by the Deutsche Forschungsgemeinschaft (DFG, German Research Foundation) within the German Excellence Strategy (Project No. 390831469).
	\appendix 
	\section{Matrix Elements}\label{sec:appA}
	In this section, we present the detailed computations for the matrix elements of the inflaton scattering with its decay products.
	\subsection*{Inflaton-Fermion Scatterings}
	\begin{figure}[!ht]
		\def\sepf{0.15}
		\centering
		\includegraphics[scale=\sepf]{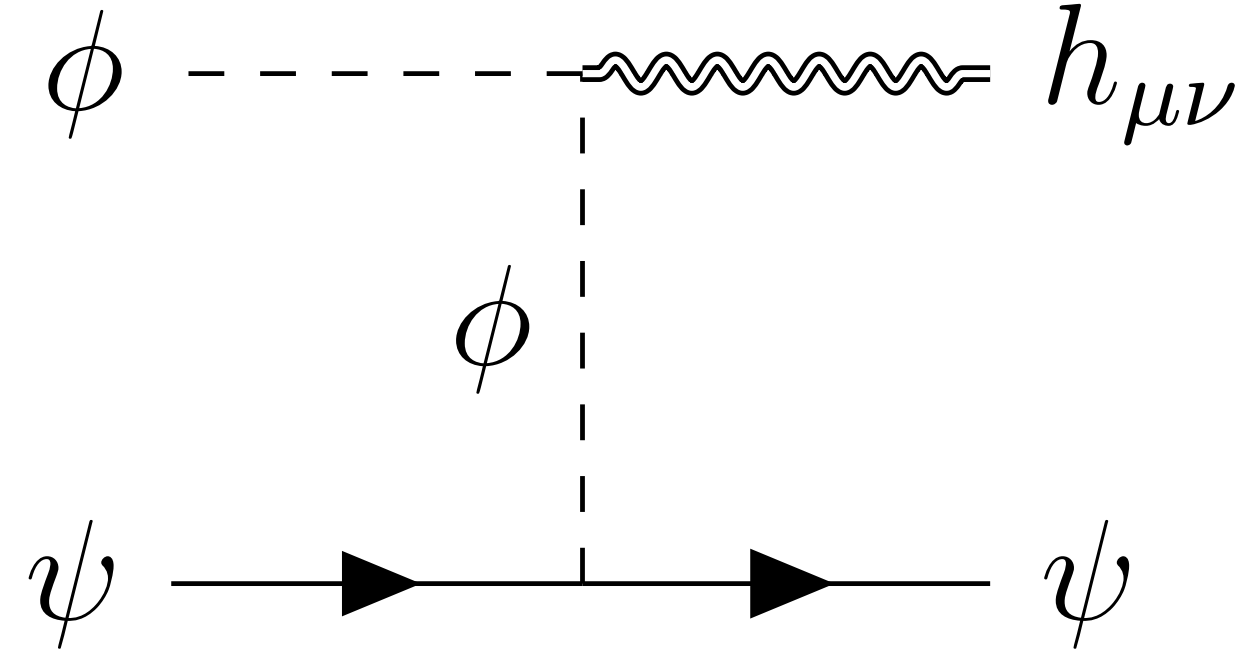}
		\includegraphics[scale=\sepf]{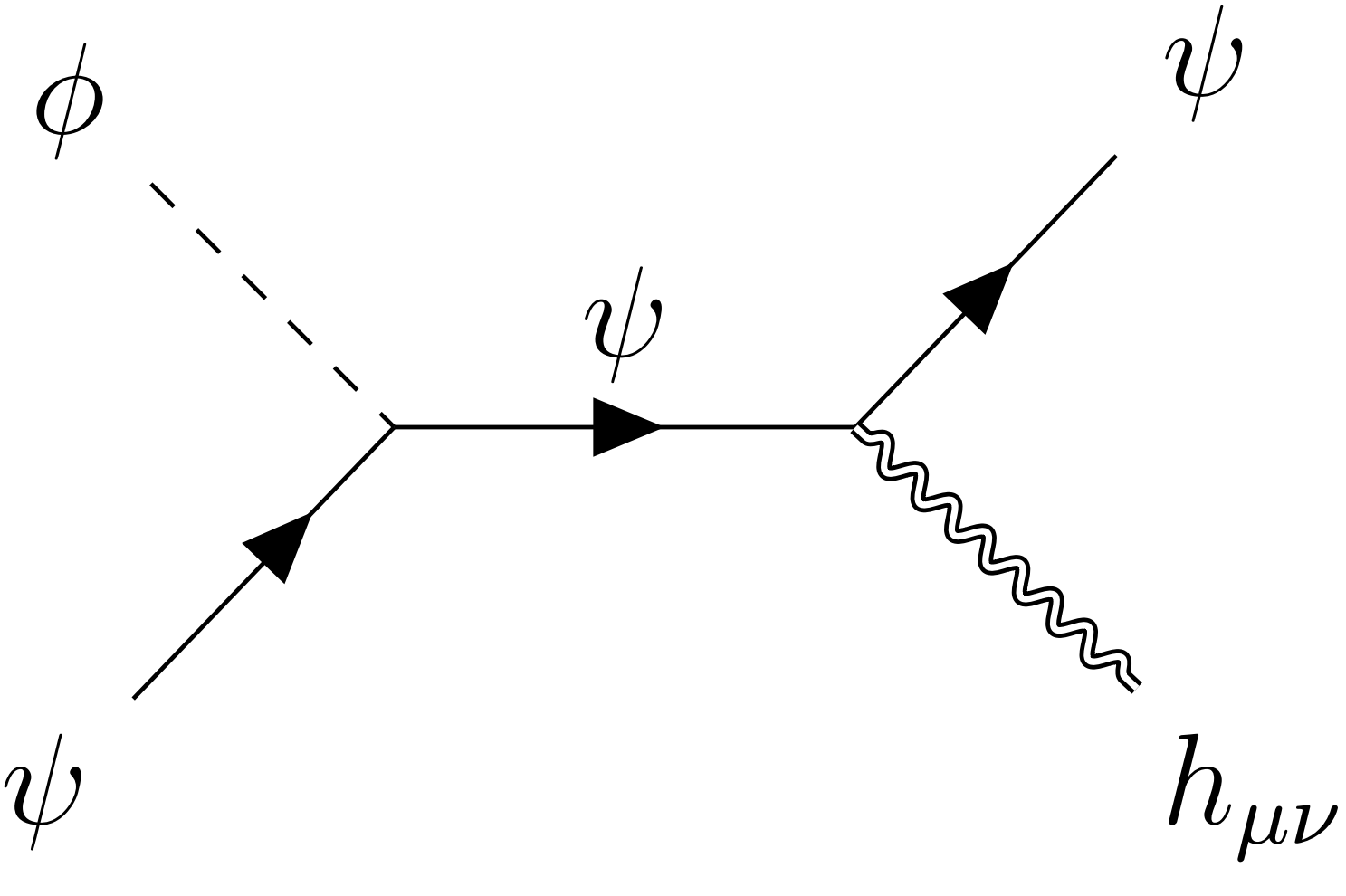}
		\includegraphics[scale=\sepf]{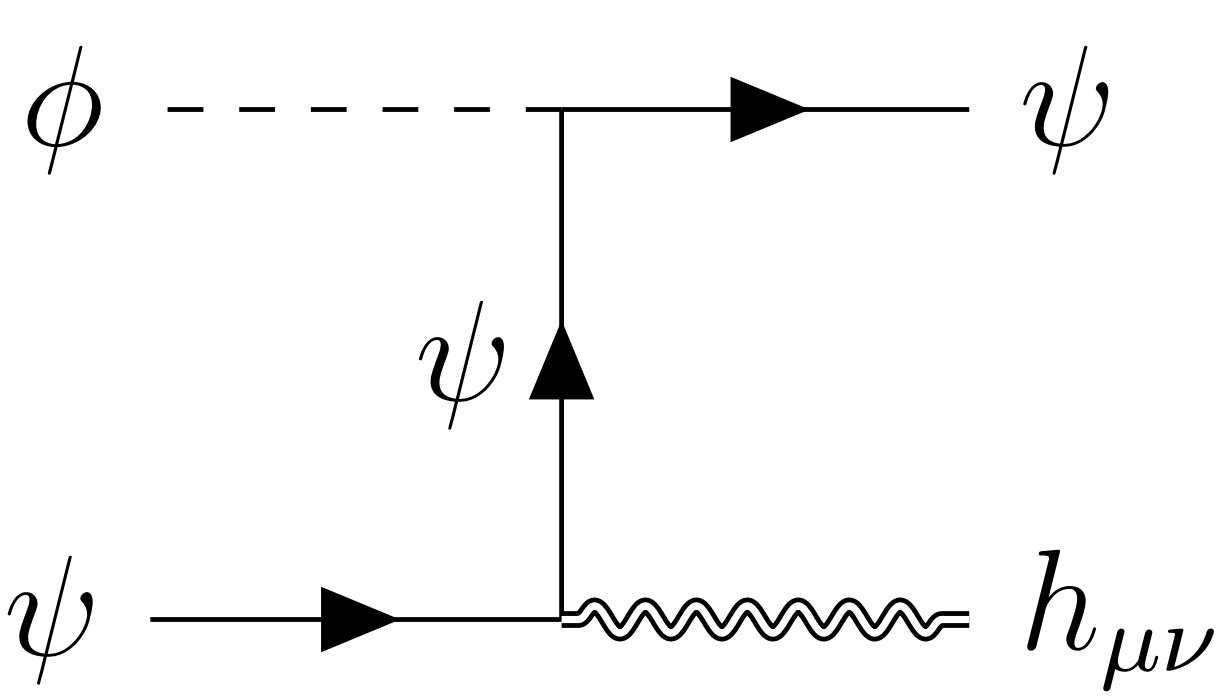}
		\includegraphics[scale=\sepf]{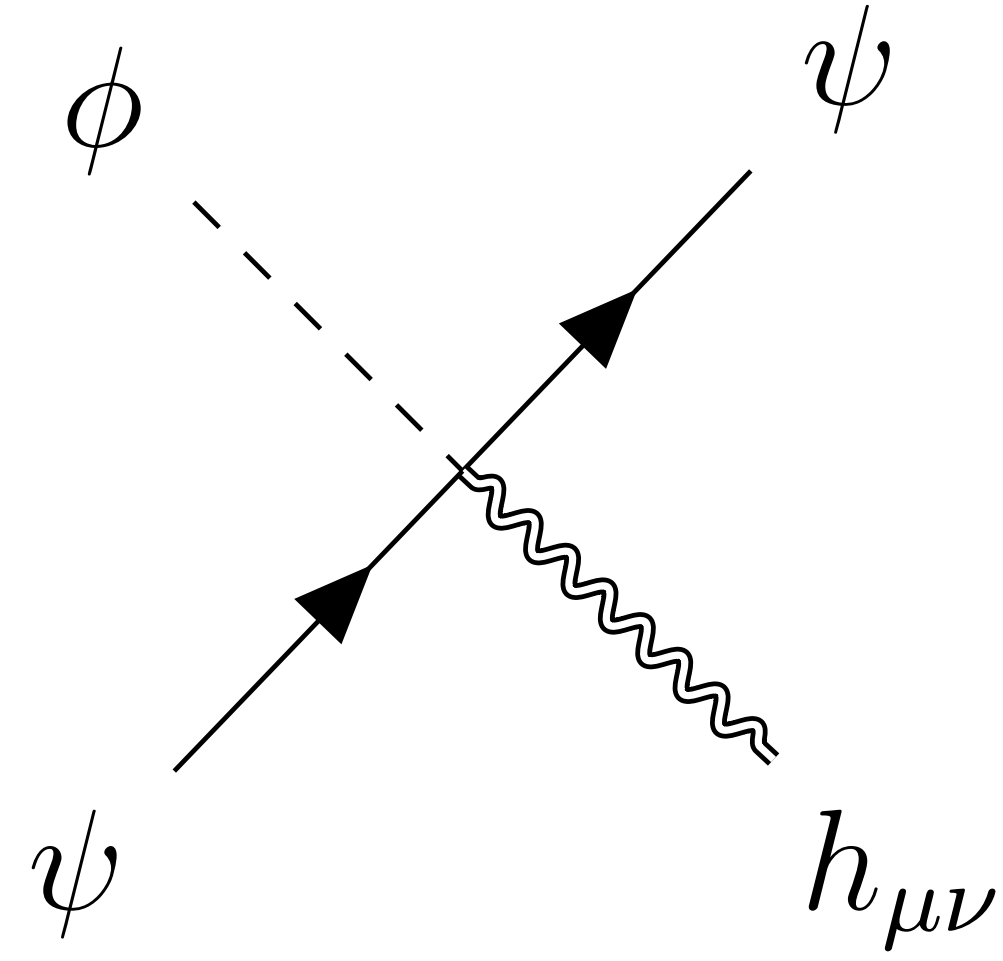}
		\caption{Graviton production via $2\to 2$ scattering between inflaton and fermionic decay product.}
		\label{fig:diagram2-2}
	\end{figure} 
	In this subsection, we study graviton production from inflaton and its fermionic decay product scattering; the corresponding Feynman diagrams are shown in Fig.~\ref{fig:diagram2-2}. We label the four-momenta as $\phi(l) + \psi(q) \to \psi(p) + h_{\mu \nu}(\omega)$. By applying the Feynman rules \cite{Choi:1994ax}, we derive the $2 \to 2$ scattering matrix elements (from left to right in Fig.~\ref{fig:diagram2-2}):
	\begin{align}
		i \mathcal{M}_1 &= \, \frac{-i y}{2 l\cdot \omega M_P}\, \left(2l_\mu\, l_\nu\right)\, \bar{u}(p) u (q)\, \epsilon^{\star\mu\nu} \,, \\
		i \mathcal{M}_2 &= \frac{i y}{2 p\cdot \omega M_P} \left[\bar{u}(p) (p_\mu \gamma_\nu) (\slashed{l}+ 2 m) u(q)\right] \epsilon^{*\mu \nu}, \\
		i \mathcal{M}_3 &= \frac{-i y}{2 q\cdot \omega M_P} \left[\bar{u}(p) (\slashed{l}- 2m)) (q_\mu \gamma_\nu)  u (q)\right] \epsilon^{*\mu \nu},\\
		i \mathcal{M}_4 &\propto \eta_{\mu \nu}\epsilon^{*\mu \nu} = 0\,,
	\end{align}
	where $\epsilon^{\mu \nu}$ denotes the graviton polarization tensor for a massless spin-2 graviton field. Note that for graviton production, an anisotropic energy momentum tensor from the source is needed. For the first diagram, the matrix element vanishes since the inflaton condensate has  vanishing three-momentum or equivalently the inflaton anisotropic energy momentum tensor $T^{ij}=0$.  The last matrix element vanishes due to the traceless condition \cite{Barman:2023ymn}. The polarization sum reads~\cite{deAquino:2011ix}
	\begin{equation}\label{eq:tensor_pol_sum}
		\sum_\text{pol} \epsilon^{\star\mu\nu} \epsilon^{\alpha\beta} = \frac12 \left(\hat{\eta}^{\mu\alpha} \hat{\eta}^{\nu\beta} + \hat{\eta}^{\mu\beta} \hat{\eta}^{\nu\alpha} - \hat{\eta}^{\mu\nu} \hat{\eta}^{\alpha \beta}\right),
	\end{equation}
	with
	\begin{equation}
		\hat{\eta}_{\mu \nu} \equiv \eta_{\mu \nu} - \frac{\omega_\mu \bar{\omega}_\nu +\bar{\omega}_\mu \omega_\nu}{\omega\cdot \bar{\omega}}\,,
	\end{equation}
	where $\omega = (E_\omega, \vec{\omega})$ and $\bar{\omega} = (E_\omega,- \vec{\omega})$.  
	
	After summing over the spin and polarization of the final states and averaging over the initial state, we find
	\begin{align}\label{eq:M2_fermion}
		|\mathcal{M}|^2 &\simeq \frac{y^2\, m_\phi^2}{2M_P^2}  \left[2 \left(\frac{E_\omega}{ m_\phi}\right)  -1\right] \left[2- 2 \left(\frac{m_\phi}{E_\omega}\right) +\left(\frac{m_\phi}{E_\omega}\right)^2 \right] \left[2 \left(\frac{E_p}{E_\omega}\right) +1 -\frac{m_\phi}{E_\omega}\right]^2\,.
	\end{align}
	with which we obtain
	\begin{align}\label{eq:ferm_2to2}
		\Gamma^{2\to 2} & \simeq n_\psi \frac{1}{32\,m_\phi  E_q\, \pi} \frac{y^2\, m_\phi^2}{2M_P^2}\left[2 \left(\frac{E_p}{E_\omega}\right) \right]^2 \nonumber \\
		&= \frac{3y^2}{4 \pi^3 } \frac{T^4}{M_P^2\,m_\phi}\,,
	\end{align} 
	where we have used equilibrium number density $ n_\psi= \frac{g_\psi\, T^3}{\pi^2}$ for massless particle with $g_\psi$ denoting the degrees of freedom. In our analysis, we have considered $g_\psi =4$, $E_q=3T$ and $\Eom =m_\phi$.  
	\subsection*{Inflaton-Boson Scatterings}
	\begin{figure}[h!]
		\def\sepf{0.15}
		\centering
		\includegraphics[scale=\sepf]{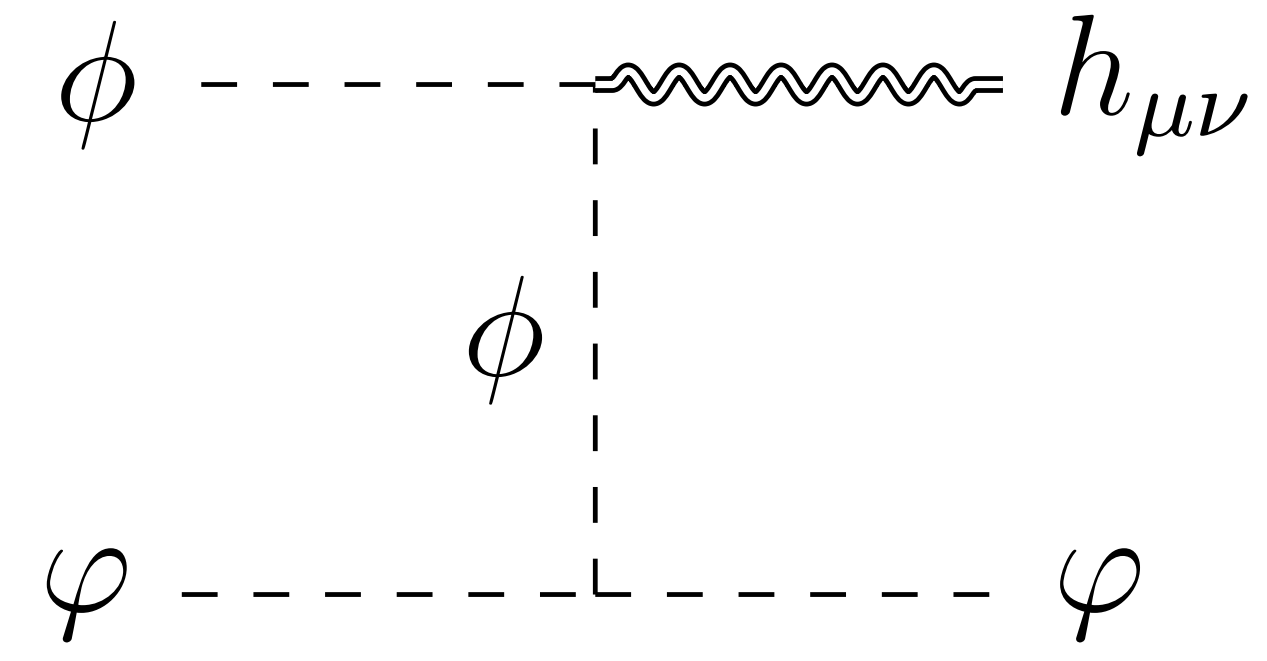}
		\includegraphics[scale=\sepf]{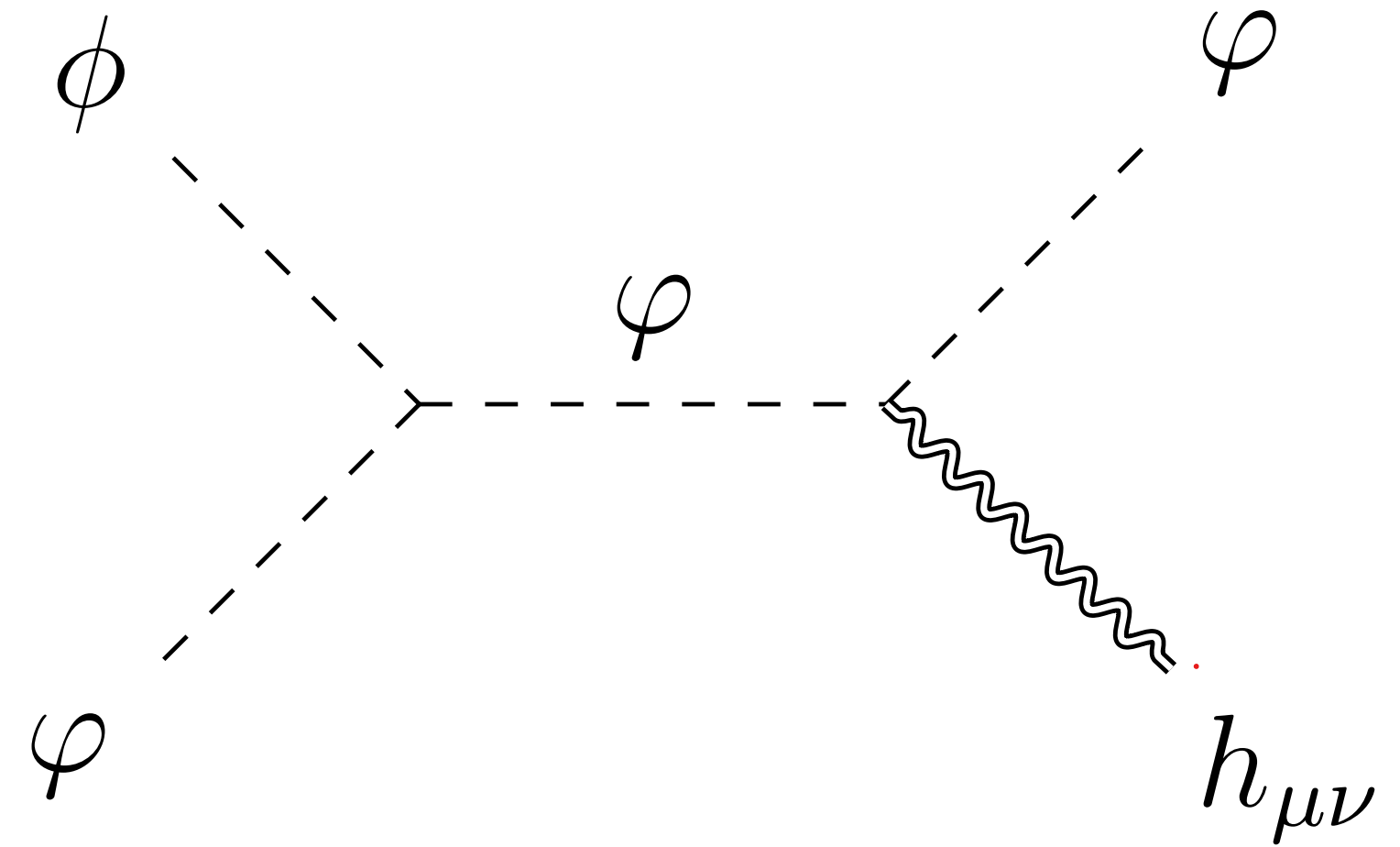}
		\includegraphics[scale=\sepf]{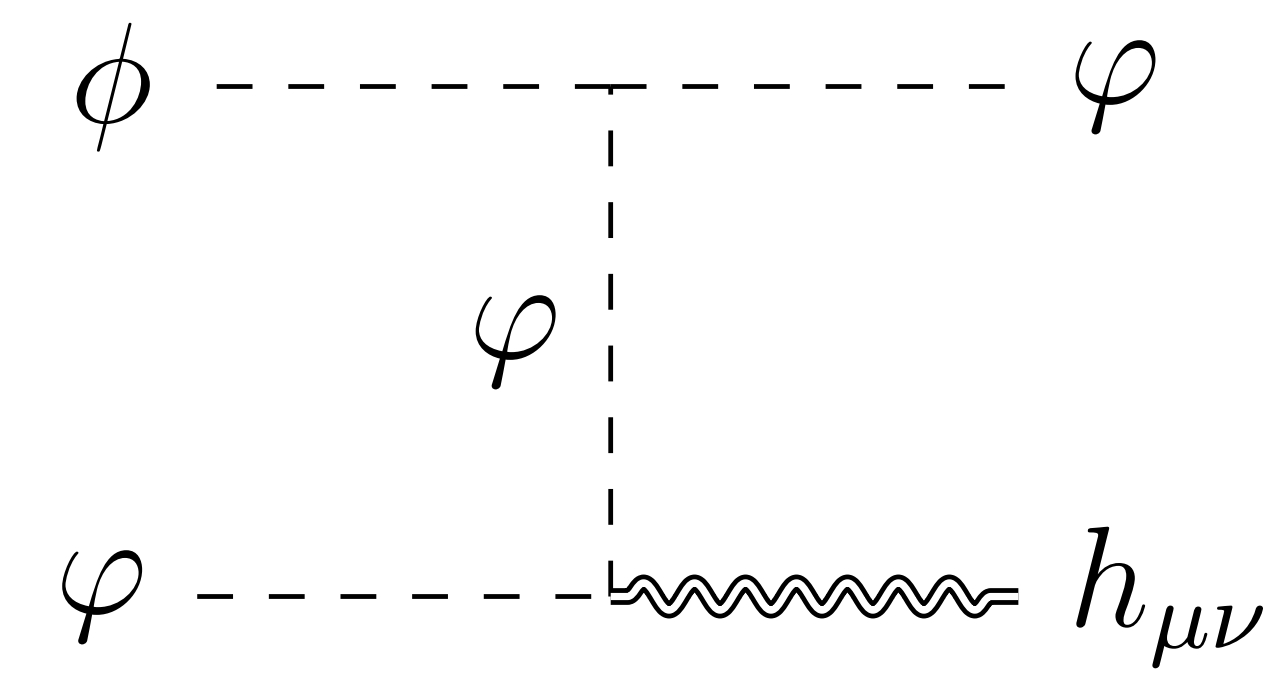}
		\includegraphics[scale=\sepf]{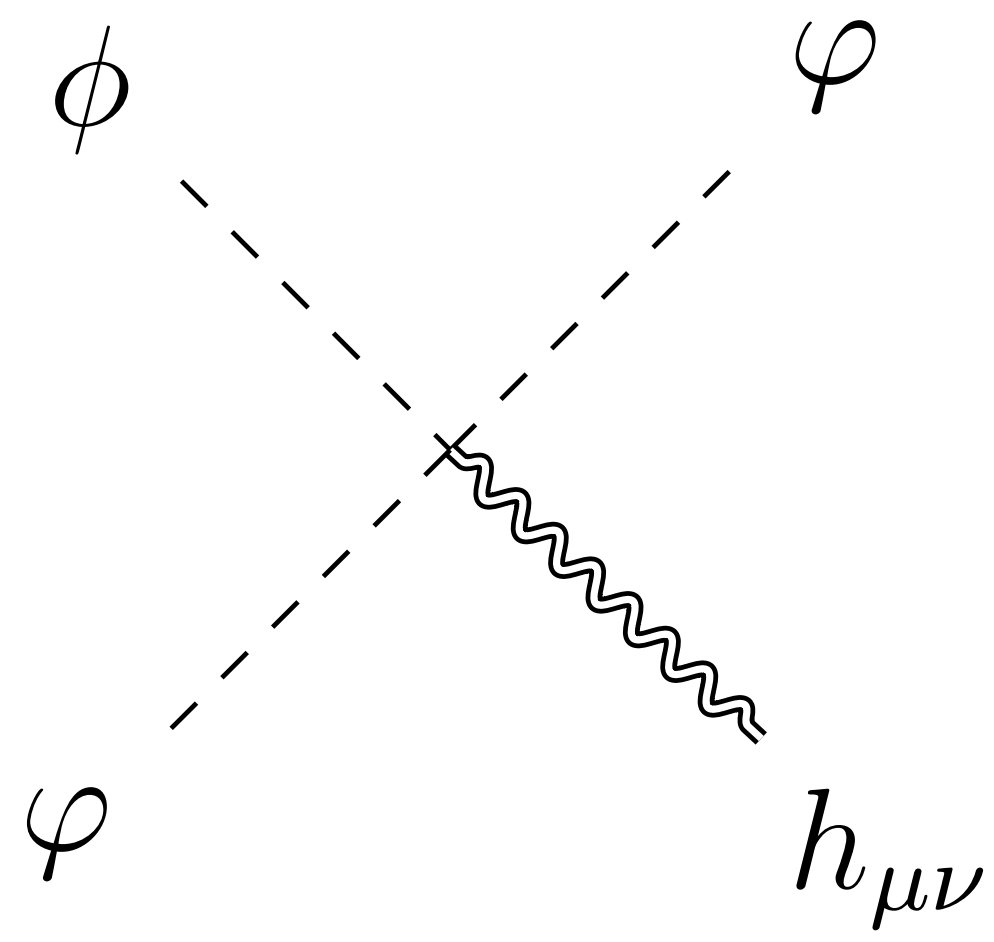}
		\caption{Graviton production via $2\to2$ scatterings between inflaton and  bosonic decay product.}
		\label{fig:diagram2-2_scalar}
	\end{figure} 
	Similarly, gravitons could also be generated from scatterings between inflaton and its bosonic decay products, as shown in Fig.~\ref{fig:diagram2-2_scalar}. We label the momenta as $\phi(l) \, \varphi(q) \to \varphi(p) \, h_{\mu \nu } (\omega)$, and the matrix elements are
	\begin{align}
		i\mathcal{M}_1 &= \frac{-i\, \mu}{M_P}\, \frac{l_\mu\, l_\nu\, \epsilon^{\star\mu\nu}}{l\cdot \omega}\,, \\
		i\mathcal{M}_2 &= \frac{i\, \mu}{M_P}\, \frac{p_\mu\, p_\nu\, \epsilon^{\star\mu\nu}}{p \cdot \omega}\,, \\
		i\mathcal{M}_3 &= \frac{-i\, \mu}{M_P}\, \frac{q_\mu\, q_\nu\, \epsilon^{\star\mu\nu}}{q\cdot \omega}\,,\\
		i\mathcal{M}_4 &\propto \eta_{\mu \nu} \epsilon^{\mu \nu} =0\,,
	\end{align} 
	where the last matrix element $\mathcal{M}_4 $ vanishes due to the traceless condition for graviton polarization tensor. The first matrix element also vanishes because the inflaton condensate behaves non-relativistically with zero three-momentum.
	
	The total squared matrix element  is after doing the polarization sum is shown to be
	\begin{align}\label{eq:M2_scalar}
		|\mathcal{M}|^2 &\simeq \frac{\mu^2}{2M_P^2}  \left[2 - \left(\frac{ m_\phi}{E_\omega}\right) 
		\right]^2 \,,
	\end{align}
	with which we find the graviton production rate:
	\begin{align}\label{eq:bos_2to2}
		\Gamma^{2\to 2} & \simeq \frac{g_\varphi\, T^3}{\pi^2} \frac{1}{32\,m_\phi  E_q\, \pi} \frac{\mu^2}{2\,M_P^2}  \nonumber \\
		&= \frac{1}{48 \pi^3 }\frac{\mu^2}{M_P^2}\frac{T^2}{m_\phi} \,.
	\end{align} 
	by considering $g_\varphi =4$, $E_q=E_p=3T$, and $\Eom =m_\phi$. Note that we have assumed scattering between the inflaton and the thermalized decay products. As mentioned earlier in the main text, scattering between the inflaton and non-thermalized $\varphi$ and $\psi$ particles can also occur before the thermal bath is developed in the pre-thermalization phase. However, such contributions to graviton production are highly suppressed due to significant entropy dilution. It is important to emphasize that we are interested in quantities at the end of reheating, where earlier graviton production leads to greater dilution.
	
	\section{Full Spectrum for Bremsstrahlung}\label{sec:full_spectrum}
\begin{figure}[!ht]
		\def\sepf{0.55}
		\centering
		\includegraphics[scale=\sepf]
		{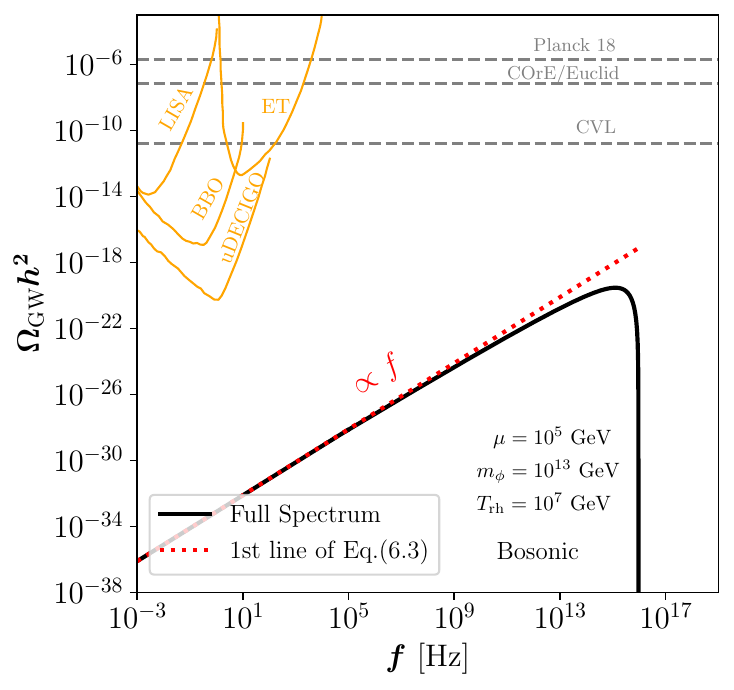}
		\includegraphics[scale=\sepf]
		{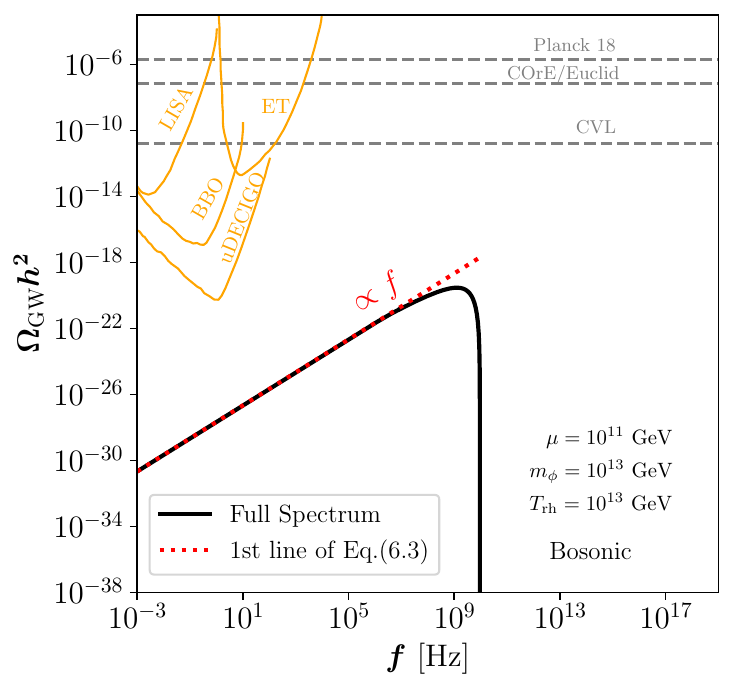}
		\includegraphics[scale=\sepf]
		{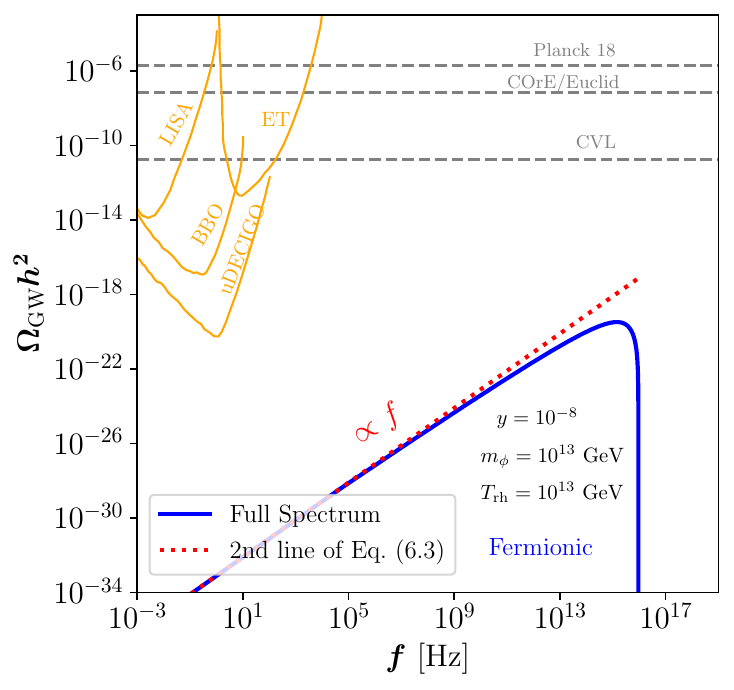}
		\includegraphics[scale=\sepf]
		{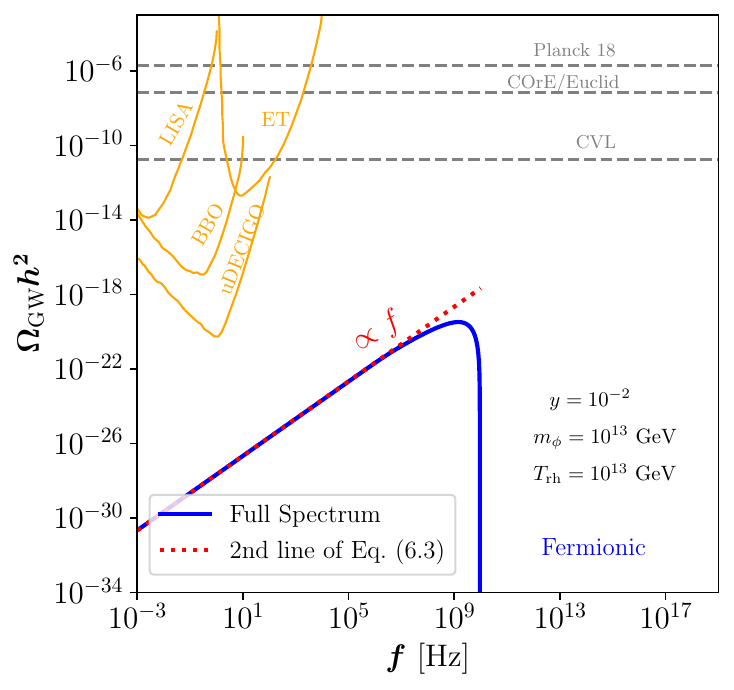}
		\caption{GW from  Bremsstrahlung as function of the couplings ($\mu$ or $y$), reheating temperature $\Trh$, inflaton mass $m_\phi$ and frequency $f$. The  solid lines are based on the full spectrum Eq.~\eqref{eq:full_bos} and the red dotted line is from Eq.~\eqref{eq:oGW2_1to3}. 
		}
		\label{fig:OGW_1to3_compare}
	\end{figure}

	Taking into account the dilution  as well as the redshift effects,  the solutions for the spectrum Eq.~\eqref{eq:Bol_rGW_diff} at the end of reheating are given  by 
	\begin{align}\label{eq:full_bos}
		\frac{d\rGW(\Trh) }{d \Eom(\Trh)} &\simeq \frac{\mu^2\, \Trh^2}{96\, \pi^2\, M_P} \sqrt{\frac{\gs}{10}} \nonumber \\ 
		& \begin{cases}
			4 \log \left(\frac{\Tmax}{\Trh}\right) - 3 \left[\frac{2\Eom(\Trh)}{m_\phi}\right] \left[\left(\frac{\Tmax}{\Trh}\right)^{8/3} - 1\right] \\
			\quad + \frac{3}{4} \left[\frac{2\Eom(\Trh)}{m_\phi}\right]^2 \left[\left(\frac{\Tmax}{\Trh}\right)^{16/3} - 1\right] \\
			\quad \text{bosonic with } 0 < \Eom(\Trh) \leq \frac{m_\phi}{2}\left(\frac{\Tmax}{\Trh}\right)^{-8/3}, \\
			\\
			\frac{3}{2} \log\left(\frac{m_\phi}{2\, \Eom(\Trh)}\right) - \frac{9}{4} + 3 \left[\frac{2\Eom(\Trh)}{m_\phi}\right] \\
			\quad - \frac{3}{4} \left[\frac{2\Eom(\Trh)}{m_\phi}\right]^2 \\
			\quad \text{bosonic with } \frac{m_\phi}{2} \left(\frac{\Tmax}{\Trh}\right)^{-8/3} \leq \Eom(\Trh) \leq \frac{m_\phi}{2},
		\end{cases}
	\end{align}
	for bosonic decay. Similarly, for fermionic decay, the solutions are 
	\begin{align}\label{eq:full_ferm}
		\frac{d\rGW(\Trh) }{d \Eom(\Trh)} &\simeq \frac{y^2\, \Trh^2\, m_\phi^2}{192 \, \pi^2\, M_P} \sqrt{\frac{\gs}{10}} \nonumber \\ 
		& \begin{cases}
			8\log \left(\frac{\Tmax}{\Trh}\right) -6 \left[\frac{2\Eom(\Trh)}{m_\phi} \right]\left[\left(\frac{\Tmax}{\Trh}\right)^{8/3} -1 \right] + \frac{9}{4}  \left[\frac{2\Eom(\Trh)}{m_\phi} \right]^2 \left[\left(\frac{\Tmax}{\Trh}\right)^{16/3} -1 \right] \\
			\quad - \frac{1}{2} \left[\frac{2\Eom(\Trh)}{m_\phi} \right]^3 \left[\left(\frac{\Tmax}{\Trh}\right)^{8} -1 \right] \quad \text{fermionic with } 0< \Eom(\Trh) \leq \frac{m_\phi}{2}\left(\frac{\Tmax}{\Trh}\right)^{-8/3}, \\ 
			3 \log\left(\frac{m_\phi}{2\, \Eom(\Trh)}\right) - \frac{17}{4} + 6 \left[\frac{2\Eom(\Trh)}{m_\phi} \right] - \frac{9}{4} \left[\frac{2\, \Eom(\Trh)}{m_\phi} \right]^2 \\
			\quad + \frac{1}{2} \left[\frac{2\Eom(\Trh)}{m_\phi} \right]^3 \quad \text{fermionic with } \frac{m_\phi}{2} \left(\frac{\Tmax}{\Trh}\right)^{-8/3}  \leq \Eom(\Trh) \leq \frac{m_\phi}{2}.
		\end{cases}
	\end{align}
	We note that when $\Eom(\Trh) = m_\phi/2$, the spectrum vanishes as expected, and the two solutions match at $\Eom(\Trh) = m_\phi/2 \left(\Tmax/\Trh\right)^{-8/3}$. In the limit $\Eom(\Trh) \ll m_\phi/2$, the differential spectrum becomes independent of $\Eom(\Trh)$, leading to Eq.~\eqref{eq:oGW2_1to3}, where the GW spectrum $\oGW \propto f$. This fits well with the spectrum in the regime $f < f_{\text{peak}}$, as shown in Fig.~\ref{fig:OGW_1to3_compare}. 
	The solid lines represent the full spectrum based on Eq.~\eqref{eq:full_bos} (upper panels) and Eq.~\eqref{eq:full_ferm} (lower panels). The model parameters considered are: $\mu = 10^{5}~\text{GeV}\,, m_\phi = 10^{13}~\text{GeV}\,, \Trh = 10^{7}~\text{GeV}$ (upper left), $\mu = 10^{11}~\text{GeV}\,, m_\phi = 10^{13}~\text{GeV}\,, \Trh = 10^{13}~\text{GeV}$ (upper right), $y = 10^{-8}\,, m_\phi = 10^{13}~\text{GeV}\,, \Trh = 10^{7}~\text{GeV}$ (lower left), and $y = 10^{-2}\,, m_\phi = 10^{13}~\text{GeV}\,, \Trh = 10^{13}~\text{GeV}$ (lower right). Note that the peak frequency $f_{\text{peak}} \simeq 9.5 \times 10^9~\text{Hz}\left(\frac{m_\phi}{\Trh}\right)$ (cf. Eq.~\eqref{eq:fpeak}) is controlled by the ratio $m_\phi/\Trh$. For a fixed inflaton mass, a lower $\Trh$ results in a higher $f_{\text{peak}}$, as seen in the left and right panels.
	%
	\section{Probing Reheating with Bremsstrahlung Gravitational Waves in a Quadratic Potential}\label{sec:probing}
	In this section, we offer a potential proposal regarding how Bremsstrahlung GWs could act as a promising avenue to probe reheating, particularly the two important parameters: $m_\phi$ and $\Trh$. As mentioned in the main text, throughout this work we have focused on a quadratic inflaton potential. For an inflaton potential steeper than quadratic, it is shown in Ref.~\cite{Barman:2023rpg} that the Bremsstrahlung GW spectrum can be enhanced with the increase in steepness of the inflaton potential during reheating, making it possible to utilize Bremsstrahlung GWs to probe the shape of the inflaton potential during reheating \cite{Barman:2023rpg}. Here, we mainly focus on a quadratic inflaton potential during reheating. The proposal and discussion presented in this section are complementary to Refs.~\cite{Barman:2023rpg, Barman:2024htg}.
	
	The couplings can be rewritten using Eq.~\eqref{eq:phiFF}, and as a result, the Bremsstrahlung GW spectrum becomes
	\begin{align}\label{eq:oGW2_1to3_2}
		\oGW^{1\to3}h^2(f)   &   \simeq  4.5  \cdot 10^{-19}  \cdot \log \left(\frac{\Tmax}{\Trh}\right)   \left( \frac{m_\phi}{10^{13}~\text{GeV}}\right)  \left( \frac{\Trh}{10^{13}~\text{GeV}}\right)   \left( \frac{f}{10^{9}~\text{Hz}}\right)\,
	\end{align}
	for  $f\lesssim f_{\text{peak}}$. At the peak, the amplitude of the spectrum is 
	\begin{align}\label{eq:oGW2_1to3_peak}
		\oGW^{1\to3}h^2(f_{\text{peak}}) \simeq  \mathcal{O} (10^{-18})    \left( \frac{m_\phi}{10^{13}~\text{GeV}}\right)^2\,,
	\end{align}
	which is dominantly controlled by the mass scale of the inflaton. For a fixed inflaton mass, it is expected that the amplitude at the peak remains almost constant. On the other hand, for smaller $m_\phi$, the peak value of the spectrum decreases.
	\begin{figure}[!ht]
		\def\sepf{0.55}
		\centering
		\includegraphics[scale=\sepf]
		{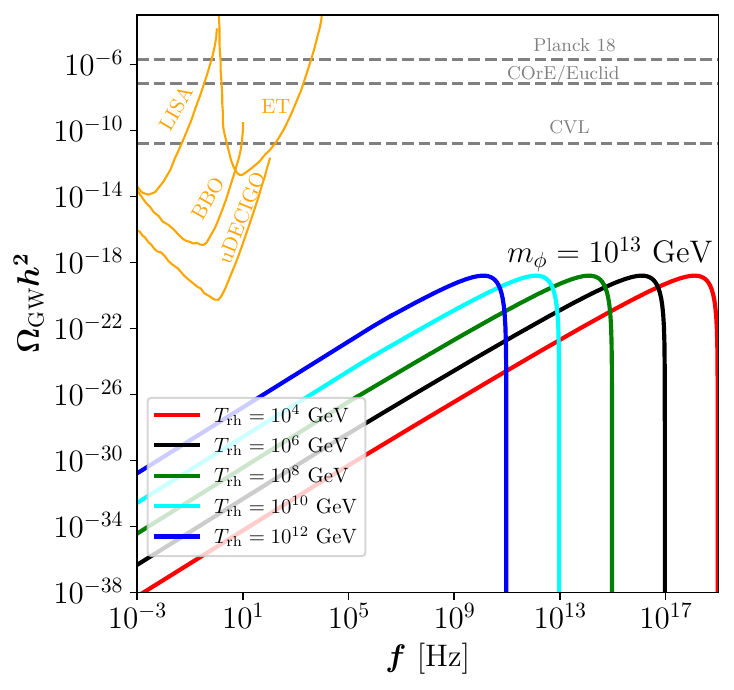}
		\includegraphics[scale=\sepf]
		{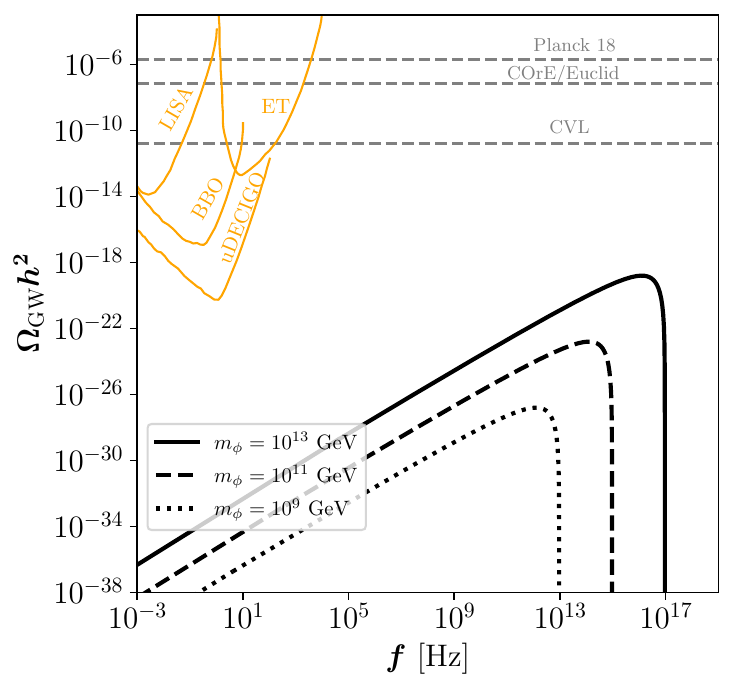}
		\caption{Dependence of Bremsstrahlung GW on $m_\phi$ and $\Trh$. In the left panel, we fix the inflaton mass, while in the right panel we vary it.}
		\label{fig:Bremsstrahlung_GW}
	\end{figure} 

	In the left panel of Fig.~\ref{fig:Bremsstrahlung_GW}, we show the Bremsstrahlung GW spectrum considering $m_\phi =10^{13}~\text{GeV}$ and $\Trh = 10^{4}~\text{GeV}$ (red solid), $\Trh = 10^{6}~\text{GeV}$ (black solid), $\Trh = 10^{8}~\text{GeV}$ (green solid), $\Trh = 10^{10}~\text{GeV}$ (cyan solid), and $\Trh = 10^{12}~\text{GeV}$ (blue solid). In the right panel, we show the spectrum for fixed $\Trh = 10^{6}~\text{GeV}$ with $m_\phi =10^{13}~\text{GeV}$ (black solid), $m_\phi =10^{11}~\text{GeV}$ (black dashed), and $m_\phi =10^{9}~\text{GeV}$ (black dotted).
	
	For fixed $m_\phi$, as expected, the amplitude at the peak is almost constant. For smaller $\Trh$, the peak occurs at higher frequencies since gravitons produced at smaller $\Trh$ receive less redshift until the present. Conversely, for larger $\Trh$, the peak frequency is smaller due to more redshifts. These features make Bremsstrahlung GWs an interesting portal to probe reheating, particularly the two important parameters $m_\phi$ and $\Trh$. 
	
	Suppose future high-frequency GW detectors see no signal at the level of $\Omega_{\text{GW}}h^2\sim 10^{-18}$ in the frequency range $10^{11}~\text{Hz} \lesssim f \lesssim 10^{13}~\text{Hz}$. This implies that either $(i)$ the inflaton mass must be $m_\phi < 10^{13}~\text{GeV}$ or $(ii) $ a reheating temperature $10^{10}~\text{GeV} \lesssim \Trh \lesssim 10^{12}~\text{GeV}$ is ruled out for $m_\phi \gtrsim 10^{13}~\text{GeV}$. This demonstrates how Bremsstrahlung GWs could potentially help probe the parameter space of reheating for a quadratic potential.  
	\bibliography{biblio}
\end{document}